\documentclass[11pt]{article}
\pdfoutput=1

\usepackage{jheppub}
\usepackage{amssymb,amsfonts,amsmath,amsthm,lineno}
\usepackage{comment}
\usepackage{enumerate}
\usepackage{mathrsfs}
\usepackage{bbold}
\usepackage{cancel}
\usepackage{tikz}
\usetikzlibrary{cd,shapes.geometric, arrows}
\usepackage{hyperref}
\usepackage{slashed}
\usepackage{color}
\usepackage{empheq}
\usepackage{soul}
\usepackage{mathdots}

\makeatletter
\newcommand{\Vast}{\bBigg@{4.75}}
\makeatother

\newcommand{\be}{\begin{equation}}
\newcommand{\ee}{\end{equation}}
\newcommand{\bea}{\begin{eqnarray}}
\newcommand{\eea}{\end{eqnarray}}

\newcommand{\CA}{\mathcal{A}}
\newcommand{\CB}{\mathcal{B}}
\newcommand{\CC}{\mathcal{C}}

\newcommand{\CF}{\mathcal{F}}

\newcommand{\CH}{\mathcal{H}}
\newcommand{\CI}{\mathcal{I}}

\newcommand{\CK}{\mathcal{K}}
\newcommand{\CL}{\mathcal{L}}
\newcommand{\CN}{\mathcal{N}}
\newcommand{\CM}{\mathcal{M}}
\newcommand{\CO}{\mathcal{O}}
\newcommand{\CP}{\mathcal{P}}

\newcommand{\CW}{\mathcal{W}}

\newcommand{\lr}{\left (}
\newcommand{\rr}{\right )}
\newcommand{\ls}{\left [}
\newcommand{\rs}{\right ]}

\newcommand\qt\tau

\newcommand{\p}{\partial}

\newcommand{\tr}{\text{tr}}

\makeatletter
\renewcommand{\@seccntformat}[1]{\csname the#1\endcsname.\,\,}
\makeatother
\allowdisplaybreaks

\DeclareMathOperator{\sgn}{sgn}

\let \savenumberline \numberline
\def \numberline#1{\savenumberline{#1.}}

\makeatletter
\def\@fpheader{\relax}
\makeatother

\def\bea{\begin{eqnarray}}
\def\eea{\end{eqnarray}}

\usetikzlibrary{shapes,arrows,chains}
\usetikzlibrary{decorations.markings}
\usetikzlibrary{decorations.pathmorphing}
\usetikzlibrary{shapes.multipart}
\tikzset{snake it/.style={decorate, decoration=snake}}

\usepackage{tikz-cd}
\usetikzlibrary{cd,decorations.markings, arrows.meta}

\newcommand{\SLR}{SL($2\,,\mathbb{R}$)}
\newcommand{\SLZ}{SL($2\,,\mathbb{Z}$)}

\newcommand{\hCz}{\hat{C}^{(0)}}

\newcommand{\hCtw}{\hat{C}^{(2)}}
\newcommand{\Ctw}{C^{(2)}}

\newcommand{\hCf}{\hat{C}^{(4)}}
\newcommand{\Cf}{C^{(4)}}

\newcommand{\rviel}[2]{\hat{E}^{}_{#1}{}^{\hat{#2}}}
\newcommand{\lviel}[2]{\tau^{}_{#1}{}^{#2}}
\newcommand{\tviel}[2]{E^{}_{#1}{}^{#2'}}
\newcommand{\invlviel}[2]{\tau^{#1}{}_{#2}}
\newcommand{\invtviel}[2]{E^{#1}{}_{#2'}}
\newcommand{\qdiv}[1]{\stackrel{(2)}{#1}}

\title{\
\vspace{1.6cm} \\
Non-Lorentzian IIB Supergravity 
from a Polynomial Realization of SL($\mathbf{2\,,\mathbb{R}}$)}
\author[a]{Eric A. Bergshoeff,}
\author[b]{Kevin T. Grosvenor,}
\author[c]{Johannes Lahnsteiner,}
\author[c]{Ziqi Yan,}
\author[d]{\\and Utku Zorba}
\emailAdd{e.a.bergshoeff@rug.nl}
\emailAdd{kevinqg1@gmail.com}
\emailAdd{johannes.lahnsteiner@su.se}
\emailAdd{ziqi.yan@su.se}
\emailAdd{utku.zorba@boun.edu.tr}
\affiliation[a]{Van Swinderen Institute, University of Groningen\\
Nijenborgh 4, 9747 AG Groningen, The Netherlands \smallskip
}
\affiliation[b]{Instituut-Lorentz, Universiteit Leiden \\
P.O. Box 9506, 2300 RA Leiden, The Netherlands \smallskip
}
\affiliation[c]{
Nordita, KTH Royal Institute of Technology and Stockholm University\\
Hannes Alfv\'{e}ns v\"{a}g 12, SE-106 91 Stockholm, Sweden \smallskip}
\affiliation[d]{
Physics Department, Bo\u{g}azi\c{c}i University\\
34342 Bebek, Istanbul, Turkey
}
\abstract{We derive the action and symmetries of the bosonic sector of non-Lorentzian IIB supergravity by taking the non-relativistic string limit.  We find that the bosonic field content is extended by a Lagrange multiplier that implements a restriction on the Ramond-Ramond fluxes. 
We show that the SL($2\,,\mathbb{R}$) transformation rules of non-Lorentzian IIB supergravity form a novel, nonlinear polynomial realization. Using classical invariant theory of polynomial equations and binary forms,
we will develop a general formalism describing the polynomial realization of SL($2\,,\mathbb{R}$) and apply it to the special case of non-Lorentzian IIB supergravity. Using the same formalism, we classify all the relevant SL($2\,,\mathbb{R}$) invariants. Invoking other bosonic symmetries, such as the local boost and dilatation symmetry, we show how the bosonic part of the non-Lorentzian IIB supergravity action is formed uniquely from these SL($2\,,\mathbb{R}$) invariants. 
This work also points towards the concept of a non-Lorentzian bootstrap, where bosonic symmetries in non-Lorentzian supergravity are used to bootstrap the bosonic dynamics in Lorentzian supergravity, without considering the fermions.}

\begin{document}

\maketitle
\vfill\eject

\section{Introduction}

Recently, there has been considerable interest in Non-Lorentzian (NL) gravity theories and their underlying NL geometries in different fields ranging from  high-energy physics, hydrodynamics to condensed matter physics. For some reviews, see \cite{Grosvenor:2021hkn,Oling:2022fft,Bergshoeff:2022eog,Bergshoeff:2022iyb,Hartong:2022lsy}. Some of these NL gravity theories have arisen in the context of non-relativistic string theory \cite{Klebanov:2000pp, Gomis:2000bd,Danielsson:2000gi}, where they describe the low-energy string dynamics provided the spacetime curvature is small. This has been supported by several beta-function calculations from the worldsheet perspective \cite{Gomis:2019zyu, Gallegos:2019icg, Yan:2019xsf, Gomis:2020fui}. In the case of non-relativistic superstring theory, one naturally considers the supersymmetric extension of the associated NL gravity theory. This leads to the concept of NL supergravity. 

It is known that the bosonic sigma model describing non-relativistic string theory can be obtained as a critical Kalb-Ramond $B$-field limit of the relativistic Polyakov sigma model, where there is a cancellation of infinities arising from the kinetic and Kalb-Ramond term \cite{Gomis:2000bd}. We refer to such a critical $B$-field limit as the \emph{non-relativistic string limit}, which has been generalized to arbitrary bosonic background fields \cite{Andringa:2012uz, Bergshoeff:2019pij, Bidussi:2021ujm} including the Ramond-Ramond (RR) potentials \cite{Ebert:2021mfu}.
Therefore, one also expects that the same is true for the corresponding supergravity theories in the target space. Indeed, it was shown that the non-relativistic string limit naturally generalizes to 10D $\CN=1$ supergravity theory, where a crucial cancellation of infinities takes place between the Einstein-Hilbert action and the terms containing the Kalb-Ramond potential \cite{Bergshoeff:2021tfn, Bergshoeff:2021bmc}.

In view of future application to NL holography, it is motivating to study the non-relativistic string limit of IIB supergravity. This formal development will provide an exciting arena for top-down constructions of holographic duals between quantum field theories (QFTs) and gravity theories enjoying NL spacetime symmetries (see, \emph{e.g.}, \cite{Gomis:2005pg}). In particular, the dual QFTs on the boundary are expected to enjoy certain Galilei-like boost symmetries. These QFTs are intimately related to interesting NL QFTs and quantum mechanical systems that are already explored in the literature, such as the fractional quantum Hall states \cite{Son:2013rqa}, the Spin Matrix Theory \cite{Harmark:2017rpg, Harmark:2018cdl, Harmark:2020vll}, and NL QFTs with SU$(1\,,n)$ spacetime symmetries \cite{Lambert:2021nol, Lambert:2021fsl}. The study of NL IIB supergravity should enable us to investigate interesting gravitational solutions, such as black hole-like objects, with potential applications to holography. 
Before exploring any of these applications in holography, it is important to first understand the formal aspects of the NL IIB theory. 

In particular, IIB supergravity famously realizes a global SL($2\,, \mathbb{R}$) symmetry \cite{Schwarz:1983wa}, which is broken to SL($2\,,\mathbb{Z}$) when lifted to IIB superstring theory \cite{Hull:1994ys, Schwarz:1995dk}. This duality maps between weakly and strongly coupled regimes in string theory and provides an important access to M-theory in 11D. A thorough understanding of the SL($2\,, \mathbb{R}$) duality in NL IIB supergravity and its SL($2\,, \mathbb{Z}$) counterpart in non-relativistic IIB superstring theory is essential for mapping out the duality web in this NL corner of string theory, which may eventually provide new insights into Matrix string/membrane theory. This is supported by the now well-established relation between the Discrete Light-Cone Quantization (DLCQ) of string theory and non-relativistic string theory, which also has a natural lift to M-theory \cite{Gomis:2000bd, Danielsson:2000gi, Bergshoeff:2018yvt, Gomis:2020izd, Yan:2021hte, Ebert:2021mfu, Bergshoeff:2022pzk}. The DLCQ is important for non-perturbative approaches to quantum chromodynamics in field theory and Matrix theory in string/M-theory \cite{Susskind:1997cw, Seiberg:1997ad, Sen:1997we, Taylor:2001vb}.

\vspace{3mm}

In this paper, we will start with the somewhat expected result that the non-relativistic string limit is also well-defined for the bosonic sector of IIB supergravity. This is the first step towards the full supergravity theory including the fermionic sector. Surprisingly, formulating the bosonic sector of NL IIB supergravity already leads us to uncover a series of intriguing novelties brought by its NL nature. We summarize the advances in this work below. 

\paragraph{Emergence of a New Field in Non-Lorentzian IIB.} We will derive an action principle of NL IIB supergravity (see Eqs.~\eqref{eq:NLIIB} and \eqref{eq:(0)S}) and show that it incorporates a new field content imposing a duality constraint on the RR fluxes (see Eqs.~\eqref{eq:o3wl2} and \eqref{eq:Omega3}). We briefly summarize the reason why this new field content emerges below. The non-relativistic string limit of the Lorentzian IIB supergravity action leads to additional divergences in the RR sector that \emph{a priori} do not cancel. However, we will show that this superficial divergence leads to constraints in the system instead of any pathology. We approach this by rewriting the divergent part via the Hubbard-Stratonovich formalism, where an auxiliary field is introduced such that the non-relativistic string limit becomes non-singular. This method is in part inspired by the worldsheet technique that leads to the Polyakov formalism of non-relativistic string theory \cite{Gomis:2000bd}. After taking the limit, the auxiliary field gives rise to the new field content in NL IIB supergravity and plays the role of a Lagrange multiplier imposing the duality constraint on the RR three- and five-form fluxes. This constraint is analogous to the self-duality condition on the RR five-form flux in Lorentzian IIB supergravity. See section~\ref{sec:nrstrlim}.

\paragraph{Branched SL($2\,,\mathbb{R}$).} We also highlight that the SL($2\,,\mathbb{R}$) transformations in NL IIB theory exhibit a branching structure. In our previous work \cite{Bergshoeff:2022iss}, we have focused on the interplay between string/brane objects in NL IIB superstring theory and the SL($2\,,\mathbb{R}$) duality, which is broken to SL($2\,,\mathbb{Z}$) due to the quantization of the $(p\,,q)$ strings \cite{Schwarz:1995dk}. Here, $p$ and $q$ are co-prime with $(p\,,q) = (1\,,0)$ representing the fundamental string and $(p\,,q)=(0,1)$ representing the D-string. We demonstrated in \cite{Bergshoeff:2022iss} that the SL($2\,,\mathbb{Z}$) transformations develop a branching structure, \emph{i.e.}, the transformation rules crucially depend on the sign of a quantity containing both the background RR zero-form $C^{(0)} (x)$ and group parameters (see Eq.~\eqref{eq:globalsl2r}). Moreover, this branching is made physically manifest in the $(p\,, q)$ space and is characterized also by the value of $C^{(0)}$\,, through the sign of the quantity $p - q \, C^{(0)} (x)$\,.
Such a branching divides the $(p\,,q)$ space into two halves and is related to the fact that there is an ambiguity in what one defines to be a string or an anti-string. Both the branches are required in order to realize the full SL($2\,,\mathbb{Z}$) group. In contrast, we will show in the current work that there is only one and the same NL IIB supergravity action realizing the full SL($2\,,\mathbb{R})$ group, which does \emph{not} exhibit any branching structure at the level of the action anymore. This is because supergravity has an extra $\mathbb{Z}_2$ symmetry, which is absent when coupled to strings. However, the SL($2\,,\mathbb{R}$) transformations themselves still have a branched structure, which we will show in this paper corresponds to different branches of a complexified dilaton field (see section~\ref{sec:bsl2zdc}).

\paragraph{Polynomial Realization of SL($2\,,\mathbb{R}$) and Invariant Theory.} We will demonstrate that the SL($2\,,\mathbb{R}$) duality of NL IIB supergravity is realized in a way that is fundamentally different from the Lorentzian case: the SL($2\,,\mathbb{R}$) transformations in NL IIB supergravity appear non-linearly as certain finite-order polynomials of a quantity $\kappa$ containing both the background scalar fields and group parameters (see Eqs.~\eqref{eq:cbcct} and \eqref{eq:unitrnsf}). This motivates us to formulate an unconventional polynomial realization \eqref{eq:gsi0matrix} of the SL($2\,,\mathbb{R}$) duality. We will show that there exists a simple connection between this polynomial realization and invariant theory in abstract algebra \cite{olver1999classical}, which deals with group actions on \emph{binary forms} that are homogeneous polynomials in two variables, or, more generically, algebraic varieties \cite{mumford1994geometric}. In particular, we will show in Eq.~\eqref{eq:tpols} that all the NL IIB data, including the new field content that we previously mentioned, and their SL($2\,,\mathbb{R}$) transformations are elegantly packaged into a quadratic and quartic binary form, which we generalize to incorporate the field strengths in IIB supergravity that are differential forms (instead of real or complex numbers). In a companion paper \cite{Ebert:2023hba}, an M-theory origin of this polynomial realization will be revealed, where the quantity $\kappa$\,, in terms of which the polynomials are formed, will have a geometric interpretation as the Galilei boost velocity on the anisotropic torus\,\footnote{The topology of this anisotropic torus is that of a pinched torus~\cite{wfmt, longpaper}.} over which non-relativistic M-theory is compactified. See section~\ref{sec:prsl2rit}. 

\paragraph{Towards a Non-Lorentzian Bootstrap.} Finally, we will apply this novel polynomial realization of global SL($2\,,\mathbb{R}$) to classify all the bosonic invariants in NL IIB supergravity (see Eqs.~\eqref{eq:sldinvs}, \eqref{eq:rtau}, and \eqref{eq:wzinv}). Focusing on the lowest-order Lagrangian terms that are quadratic in spacetime derivatives, and combined with the other bosonic symmetries that, most importantly, contain the a local Galilei-like boost symmetry and an emergent dilatation symmetry (see section~\ref{sec:symmetries}), we show that the resulting effective action in Eq.~\eqref{eq:seftf} only has a single free coupling that is the gravitational constant. Fascinatingly, this bosonic effective action precisely reproduces the NL IIB action we derive from the non-relativistic string limit. In this sense, we are able to build the action of the bosonic sector of NL IIB supergravity from the polynomial realization of SL($2\,,\mathbb{R}$), without explicitly referring to fermions or supersymmetry. Moreover, we will argue that the bosonic part of the Lorentzian IIB supergravity including the self-duality condition can be recovered by tracing back the limiting procedure. This success opens up the possibility of constructing higher-order terms in Lorentzian IIB supergravity from the NL corner coupled to non-relativistic strings using bosonic symmetries, which is essentially an ``NL bootstrap" in the sense that we use the smaller NL theory to constrain the couplings in the parent Lorentzian theory. See section~\ref{sec:EFT}. 

\vspace{3mm}

This work is organized as follows. In section~\ref{sec:nliibs}, we derive the bosonic action \eqref{eq:NLIIB} and symmetries of NL IIB supergravity by taking the non-relativistic string limit of Lorentzian IIB supergravity. In section~\ref{sec:prsl2rit}, we first develop a general polynomial realization \eqref{eq:gsi0matrix} of \SLR\ in NL theories. Then, in section~\ref{sec:gitbf}, we generalize classical invariant theory and study its application to the polynomial realization of SL($2\,,\mathbb{R}$). In preparation for the application to NL IIB theory, we classify the quadratic invariants in section~\ref{sec:cqi}. Finally, in section~\ref{sec:anl2bs}, we apply the formalism developed in section~\ref{sec:prsl2rit} to build the bosonic sector of NL IIB supergravity from symmetry principles. We first recast the NL IIB data into two binary forms in Eq.~\eqref{eq:tpols} and derive their SL($2\,,\mathbb{R}$) transformations in Eq.~\eqref{eq:unitrnsf} using the polynomial realization. Then, in section~\ref{sec:sl2rinl}, we classify all possible \SLR\ invariants that are quadratic in spacetime derivatives. Finally, we demonstrate how the NL IIB gravity action uniquely follows as a bosonic effective field theory in Section~\eqref{sec:EFT} using all the bosonic symmetries including the \SLR\, and propose the concept of NL bootstrap. We give our outlook in section~\ref{sec:outlook}. Three appendices have been added to explain some of the details of our calculations.

\section{Non-Relativistic String Limit in IIB Supergravity} \label{sec:nliibs}

In this section, we derive the Non-Lorentzian (NL) IIB supergravity action from the non-relativistic string limit of the well-known Lorentzian IIB supergravity action. We will see that this leads us to extend the field content by an additional five-form field. This field appears as a Lagrange multiplier imposing a constraint on the RR fluxes. What is more, we will study the symmetries of the new NL action, highlighting the unconventional realization of SL($2\,,\mathbb{R}$) symmetry in the NL system. We will discuss the subtleties associated with the branched structure of the SL($2\,,\mathbb{R}$) transformations, where, unlike in the Lorentzian case, important sign changes are present. 

\subsection{Review of Lorentzian IIB Supergravity}

We begin with a review of the manifestly SL($2\,,\mathbb{R}$)-invariant\,\footnote{In type IIB superstring theory, the continuous \SLR\ symmetry group of IIB supergravity becomes quantized and is broken to \SLZ\ \cite{Schwarz:1995dk}.} formulation of Lorentzian IIB supergravity theory, using the Einstein frame. In this work, we focus only on the bosonic sector by setting all of the fermions to zero. In the following, we use hatted notation in Lorentzian supergravity and unhatted notation in non-Lorentzian supergravity. The action is given by \cite{Schwarz:1995dk}
\begin{align} \label{eq:relaction}
\begin{split}
    \hat{S} = \frac{1}{16\pi G^{}_\text{N}} & \int d^{10} x \, \hat{E} \, \biggl[ \hat{R} + \frac{1}{4} \, \tr \Bigl( \partial_{\mu} \hat{\CM} \, \partial^{\mu} \hat{\CM}^{-1} \Bigr) - \frac{1}{12} \, \hat{\CH}_{\mu\nu\rho}^{\intercal} \, \hat{\CM} \, \hat{\CH}^{\mu\nu\rho} \biggr] \\[4pt]
    + \frac{1}{16\pi G^{}_\text{N}} & \int \frac{1}{4} \, \biggl( \hat{F}^{(5)}\wedge \star \hat{F}^{(5)} - \hat{\CC}^{(4)} \wedge \hat{\CH}^{(3)}{}^\intercal_{\phantom{I}} \wedge \epsilon \, \hat{\CH}^{(3)} \biggr) \,,
\end{split}
\end{align}
where $G^{}_\text{N}$ is Newton's gravitational constant and $\epsilon$ is the 2D Levi-Civita symbol.
Furthermore, $\rviel{\mu}{A}$ is the Vielbein field in the Einstein frame, where $\mu = 0, \ldots, 9$ are the spacetime indices and $\hat{A} = 0, \ldots , 9$ are the frame indices. 
In addition, $\hat{E} = \det \rviel{\mu}{A}$ and $\hat{R}$ is the Ricci scalar. We also introduce the dilaton field $\hat{\Phi}$, the Kalb-Ramond two-form field $\hat{B}^{(2)}$, and the RR zero-, two-, and four-form potentials $\hCz$, $\hCtw$, and $\hat{C}^{(4)}$. We define these RR potentials to be the ones that naturally couple to the D-branes via the Chern-Simons action but in general transform non-trivially under the SL($2\,,\mathbb{R}$) group action. In supergravity, it is convenient to define a calligraphic four-form field $\CC^{(4)}$ to be
\begin{equation}
    \hat{\CC}^{(4)} = \hCf + \frac{1}{2} \hat{B}^{(2)} \wedge \hCtw\,,
\end{equation}
which is invariant under the SL($2\,,\mathbb{R}$) group action. 
The fields $\hat{\CM}$, $\hat{\CH}$, and $\hat{F}^{(5)}$ in Eq.~\eqref{eq:relaction} are derived quantities given in terms of the fundamental fields as follows:
\begin{subequations}
\begin{align}
    \hat{\CM} &= e^{\hat{\Phi}} 
\begin{pmatrix}
    \bigl( \hCz \bigr)^2 + e^{-2 \hat{\Phi}} \phantom{oo} & \hCz \\[4pt]
    \hCz \phantom{o} & 1
\end{pmatrix}\,, \\
    \hat{\CH}^{(3)} &= d \hat{\Sigma}^{(2)} \qquad {\rm where} \qquad  \hat{\Sigma}^{(2)} = \binom{\hat{B}^{(2)}}{\hat{C}^{(2)}}\,, \\[4pt]
    \hat{F}^{(5)} &= d\hat{\CC}^{(4)} + \frac{1}{2} \, \hat{\CH}^{(3)}{}^\intercal_{\phantom{I}} \wedge \epsilon \, \hat{\Sigma}^{(2)}\,.
\end{align}
\end{subequations}
A nontrivial feature of the action \eqref{eq:relaction} is that it does not give the correct equations of motion unless we impose the following self-duality condition by hand \emph{after} varying the action:
\begin{equation} \label{eq:sdcf5}
    \hat{F}^{(5)} = \star \hat{F}^{(5)}\,.
\end{equation}
For this reason, Eq.~\eqref{eq:relaction} is sometimes called a \emph{pseudo-action}.

Under an SL($2\,,\mathbb{R}$) transformation generated by the matrix
\begin{equation}\label{eq:SLParameter}
    \Lambda = 
\left( 
\begin{array}{cc}
    \alpha &\quad \beta \\
    \gamma &\quad \delta
\end{array}
\right),
    \qquad \alpha\,, \beta\,, \gamma\,, \delta \in \mathbb{R} \,, \qquad \alpha \, \delta - \beta \, \gamma = 1\,,
\end{equation}
the various fields transform as follows:
\begin{subequations}\label{eq:RSL(2)}
\begin{align}
    \hat{E}_{\mu}{}^{\hat{A}} & \rightarrow \rviel{\mu}{A}\,, 
        &
    \hat{\CC}^{(4)} & \rightarrow \hat{\CC}^{(4)}\,, 
        &
    \hat{F}^{(5)} &\rightarrow \hat{F}^{(5)}\,, 
\end{align}
and
\begin{align}
    \hat{\CH}^{(3)} &\rightarrow \bigl( \Lambda^{-1} \bigr)^{\intercal} \, \hat{\CH}^{(3)} \,,         &
    \hat{\CM} &\rightarrow \Lambda \, \hat{\CM} \, \Lambda^{\intercal} \,,
\end{align}
\end{subequations}
from which it is easy to see that the action \eqref{eq:relaction} is indeed SL($2\,,\mathbb{R}$)-invariant. In fact, each individual term in Eq.~\eqref{eq:relaction} is manifestly SL($2\,,\mathbb{R}$) invariant on its own.

The IIB fields we have introduced above transform under coordinate transformations $x \rightarrow x'(x)$ in the usual way (\emph{i.e.}, as tensors): each spacetime index being multiplied by the Jacobian for the change of coordinates or its inverse. Similarly, under the local Lorentz transformations, each frame space index is multiplied by the Lorentz transformation matrix.

The action is also gauge-invariant under the infinitesimal transformations generated by a one-form $\hat{\xi}^{(1)}$, and a collection of $p$-forms $\hat{\zeta}^{(p)}$:
\begin{align} \label{eq:gtrnsfbc}
    \delta \hat{B}^{(2)} &= d \hat{\xi}^{(1)}\,, &%
    \delta \hat{C}^{(q)} &= d \hat{\zeta}^{(q-1)} + \hat{H}^{(3)} \wedge \hat{\zeta}^{(q-3)}\,,
\end{align}
where $q = 0, 2, 4$\,. We take it for granted that $\hat{\zeta}^{(p)}$ vanishes for negative $p$\,. Moreover, $\hat{H}^{(3)}$ is the Kalb-Ramond field strength,
\begin{equation}
    \hat{H}^{(3)} = d \hat{B}^{(2)}\,.
\end{equation}

We emphasize that the calligraphic $\hat{\CC}^{(4)}$ is indeed invariant under SL($2\,,\mathbb{R}$) in Eq.~\eqref{eq:RSL(2)}, which is why it is convenient to introduce it in the first place. On the other hand, in order to have a nice unified notation for the gauge transformations, we have written them using $\hCf$ instead of $\hat{\CC}^{(4)}$ in Eq.~\eqref{eq:gtrnsfbc}. Thus, we will keep both for the convenience of discussions.

\subsection{Non-Relativistic String Limit}\label{sec:nrstrlim}

We now discuss how to derive NL IIB supergravity from Lorentzian IIB supergravity that we have discussed above. Analogous to how Lorentzian supergravity arises from relativistic string theory, NL supergravity arises from non-relativistic string theory, which is a unitary and ultra-violet complete string theory that has been studied from first principles using worldsheet techniques \cite{Gomis:2000bd} (see \cite{Oling:2022fft} for a recent review). It is also known that this string theory can be embedded within the framework of relativistic string theory via the so-called \emph{non-relativistic string limit}. 
We now perform the non-relativistic string limit of the bosonic sector of the Lorentzian IIB supergravity action following the prescriptions in \cite{Bergshoeff:2019pij, Ebert:2021mfu}. The resulting NL supergravity naturally couples to non-relativistic string theory. 

In practice, we will first reparametrize the relativistic IIB supergravity action \eqref{eq:relaction} by performing an invertible redefinition of fields that involves a dimensionless parameter $\omega$. For the RR $q$-form potentials, we use the results of \cite{Ebert:2021mfu}, where the probe D-branes were used to derive the appropriate reparametrizations. Then, we will discuss how to take a consistent $\omega \rightarrow \infty$ limit of the IIB action. The consistent non-relativistic limit of the Neveu-Schwarz action, which is a truncation of IIB supergravity in which the RR fields are set to zero, was previously obtained in \cite{Bergshoeff:2021bmc}.

We begin by splitting the range of the frame index $\hat{A} = 0\,, \ldots , 9$ into a 2D \emph{longitudinal} sector with $A = 0\,,1$ and an 8D \emph{transverse} sector with $A' = 2\,, \ldots , 9$\,. This leads to a natural split of the Lorentzian Vielbein $\rviel{\mu}{A}$ into longitudinal $\lviel{\mu}{A}$ and transverse $\tviel{\mu}{A}$ components. Then, we make the $\omega$-dependent reparametrizations of the NS-NS fields,\,\footnote{The powers of $\omega$ here are appropriate for the Einstein frame. The string frame Vielbeine are simply $e^{\Phi / 4} \lviel{\mu}{A}$ and $e^{\Phi / 4} \tviel{\mu}{A}$. The other fields are identical in both frames.}
\begin{subequations} \label{eq:NLredefinitions}
\begin{align}
    \hat{E}^{}_\mu{}^A & = \omega^{3/4} \, \tau_\mu{}^A\,, 
        &%
    \hat{\Phi} &= \Phi + \ln \omega\,, \\[4pt]
    \hat{E}^{}_\mu{}^{A'} & = \omega^{-1/4} \, E^{}_\mu{}^{A'}\,,
        &
    \hat{B}^{(2)} & = - \omega^2 \, e^{\Phi/2} \, \ell^{(2)} + B^{(2)}\,,
\end{align}
together with the reparametrizations of the Ramond-Ramond potentials,
\be \label{eq:rpcq}
    \hat{C}^{(q)} = \omega^2 \, e^{\Phi/2} \, \ell^{(2)} \wedge C^{(q-2)} + C^{(q)}\,,
\ee
\end{subequations}
where $\ell^{(2)}$ is the worldsheet volume 2-form with components defined by
\begin{equation}
    \ell_{\mu\nu} = \lviel{\mu}{A} \lviel{\nu}{B} \epsilon_{AB}\,.
\end{equation}
Note that the $\omega^2$ term in Eq.~\eqref{eq:rpcq} is understood to vanish when $q = 0$\,.

Before performing the $\omega \rightarrow \infty$ limit in the Lorentzian IIB action \eqref{eq:relaction}, we first collect some necessary definitions. We define the inverse Vielbeine fields via the orthonormality condition of $\hat{E}_{\mu}{}^{\hat{A}}$\,, which now breaks up into the following conditions: 
\begin{equation}
\begin{aligned}
    \lviel{\mu}{A} \, \invlviel{\mu}{B} &= \delta_{B}^{A}\,, \qquad &%
    \lviel{\mu}{A} \, \invlviel{\nu}{A} + \tviel{\mu}{A} \invtviel{\nu}{A} &= \delta_{\mu}^{\nu}\,, \\[4pt]
    \tviel{\mu}{A} \invtviel{\mu}{B} &= \delta_{B'}^{A'}\,, \qquad &%
    \lviel{\mu}{A} \, \invtviel{\mu}{A} = 
    \invlviel{\mu}{A} \, \tviel{\mu}{A} & = 0\,.
\end{aligned}
\end{equation}
In the rest of the paper, we will often replace spacetime indices with frame indices. This means that the spacetime index has been contracted with a Vielbein component, either $\lviel{\mu}{A}$ or $\tviel{\mu}{A}$ or their inverses depending on whether the frame index is longitudinal or transverse and raised or lowered. 
For example, 
\begin{align}
    F^{A} &= \lviel{\mu}{A} F^{\mu}\,, &%
    G_{A'} &= E^\mu{}_{A'} G_{\mu}\,.
\end{align}
We define the determinant $E$\,, $q$-form field strengths $F^{(q)}$\,, and the four-form field $\mathcal{C}^{(4)}$ in the non-Lorentzian IIB theory after the $\omega \rightarrow \infty$ limit to be
\begin{subequations} \label{eq:nlfs}
\begin{align}
        E & = \det \bigl( \lviel{\mu}{A} , \tviel{\mu}{A} \bigr)\,, 
        & %
        \CC^{(4)} &= \Cf + \frac{1}{2} B^{(2)} \wedge \Ctw\,,\label{eq:defCC4}\\[4pt]
        H^{(3)} &= d B^{(2)}\,,
        &
        F^{(q)} &= d C^{(q-1)} + C^{(q-3)} \wedge d B^{(2)}\,, \label{eq:defEdefF}
\end{align}
\end{subequations}
for $q = 1, 3, 5$ and where it is understood that $C^{(q)} = 0$ for negative $q$.

We also define the quantity
\begin{equation} \label{eq:ttdef}
    \tau^{}_{\mu\nu}{}^{A} = \partial_{[ \mu}^{\phantom{A}} \tau^{}_{\nu ]}{}^{A}
\end{equation}
for convenience. This enters into $d \ell^{(2)}$ as follows:
\begin{equation}
    d \ell^{(2)} = \epsilon^{}_{AB} \, \tau^{}_{\mu\nu}{}^{A} \, \lviel{\rho}{B} \, dx^{\mu} \wedge dx^{\nu} \wedge dx^{\rho}\,.
\end{equation}
In the rest of the paper, it is useful to remember that the wedge product of two or more $\ell^{(2)}$'s or an $\ell^{(2)}$ with $d \ell^{(2)}$ vanishes. 
\emph{A priori}, we will not impose any condition on $\tau^{}_{\mu\nu}{}^A$ (or $d\ell^{(2)}$), though, like all fields, it will be constrained by the equations of motion. Nevertheless, one can see that the condition that $d \ell^{(2)} = 0$ is the worldsheet analog of ``absolute time'' ($d \tau = 0$) for the particle \cite{Bergshoeff:2022fzb}. Moreover, extra constraints on $\tau^{}_{\mu\nu}{}^A$ usually arise once the supersymmetry transformations are concerned \cite{Bergshoeff:2021bmc} or further bosonic symmetries are imposed \cite{Andringa:2012uz, Bergshoeff:2018yvt, Yan:2021lbe}. In order to decide the necessary supersymmetric constraints on $\tau^{}_{\mu\nu}{}^A$\,, a thorough analysis of the full NL IIB supergravity including the fermionic sector has to be performed, which is beyond the scope of this paper.

Plugging the above redefinitions Eq.~\eqref{eq:NLredefinitions} into the IIB supergravity action \eqref{eq:relaction} results in an expansion of the action with respect to large $\omega$ of the form
\begin{align} \label{eq:somegaexp}
    \hat{S} = \omega^2 \qdiv{S} + \stackrel{(0)}{S} + \, \omega^{-2} \hspace{-0.1cm} \stackrel{(-2)}{S} + O(\omega^{-4})\,.
\end{align}
Ultimately, we are interested in the finite part $\stackrel{(0)}{S}$, but first we must deal with the $\omega^2$ terms in $\qdiv{S}$\,, which lead to a superficial quadratic divergence in the infinite $\omega$ limit. We will show that this superficial divergence can be tamed into a finite contribution in the resulting NL IIB supergravity action.

\paragraph{Quadratic Divergence.} The $\omega^2$ terms in Eq.~\eqref{eq:somegaexp} are given by (up to boundary terms)\,\footnote{We remind the reader that the Hodge star operation maps $p$-forms to $(10-p)$-forms and is defined via its action on an orthonormal basis $e^a$\,, $a = 0\,, 1\,, \cdots, 9$ as follows \cite{Freedman:2012zz}:
\begin{align}
    \star\Bigl(e^{a^{}_1}\wedge \cdots \wedge e^{a^{}_p}\Bigr) = \frac{1}{(10-p)!} \, e^{a^{}_{p+1}} \wedge \cdots\wedge e^{a^{}_{10}} \, \epsilon^{}_{a^{}_{p+1} \cdots a^{}_{10}}{}^{a^{}_1 \cdots a^{}_p}\,.
\end{align}
This definition is independent of the local symmetries and is thus valid for both the Lorentz structure with frame fields $\{e^a\} = \{\hat{E}^{\hat A}\}$ and the Galilei structure with frame fields $\{e^a\} = \{\tau^A,\,E^{A'}\}$.
}
\begin{align} \label{eq:S2orig}
    \omega^2 \stackrel{(2)}{S} 
    &= \frac{\omega^2}{16 \pi \, G^{}_\text{N}} \int \bigl( \Omega^{(3)} \wedge \ell^{(2)} \bigr) \wedge \star \bigl( \Omega^{(3)} \wedge \ell^{(2)} \bigr)\,,
\end{align}
where we have defined the three-form
\begin{align} \label{eq:Omega3}
    \Omega^{(3)} = \frac{1}{2} \Bigl[ \star \bigl( F^{(5)} \wedge \ell^{(2)} \bigr) - e^{\Phi /2} F^{(3)} \Bigr]\,.
\end{align}
See Eq.~\eqref{eq:qdiv} in appendix~\ref{app:eiibsa} for details of the computation of Eq.~\eqref{eq:S2orig}.  
Now, we can perform a Hubbard-Stratonovich transformation by introducing an auxiliary five-form field $\mathcal{A}^{(5)}$\,:
\begin{align} \label{eq:HS}
    \omega^2 \! \stackrel{(2)}{S} & \rightarrow \frac{1}{16 \pi G^{}_\text{N}} \int \biggl[ \mathcal{A}^{(5)} \wedge \star \bigl( \Omega^{(3)} \wedge \ell^{(2)} \bigr) - \frac{1}{4 \omega^2} \, \mathcal{A}^{(5)} \wedge \star \mathcal{A}^{(5)} \biggr]\,.
\end{align}
Using this auxiliary field trick, we have traded the $O( \omega^2 )$ term with an $O ( \omega^0 )$ and an $O ( \omega^{-2} )$ term, thereby effectively removing the divergent $O( \omega^2 )$ term altogether.
The equation of motion from  varying $\CA^{(5)}$ in Eq.~\eqref{eq:HS} is
\begin{equation}\label{eq:A5solution}
    \CA^{(5)} = 2 \, \omega^2 \, \Omega^{(3)} \wedge \ell^{(2)} \,.
\end{equation}
Indeed, if we plug this back into Eq.~\eqref{eq:HS}, we get back the original quadratic divergence Eq.~\eqref{eq:S2orig}.

If we now take the $\omega \rightarrow \infty$ limit in Eq.~\eqref{eq:HS}, then there is no longer any quadratic divergence and $\CA^{(5)}$ becomes a Lagrange multiplier imposing the constraint
\begin{equation} \label{eq:o3wl2}
     \Omega^{(3)} \wedge \ell^{(2)} = 0\,.
\end{equation}
In components of the frame indices, Eq.~\eqref{eq:o3wl2} is equivalent to $\Omega_{A'B'C'} = 0$\,. Using Eq.~\eqref{eq:Omega3}, we find
\begin{equation}\label{eq:transselfdual}
    F_{A_1' \cdots A_5'} = - \frac{e^{\Phi/2}}{3!} \, \epsilon_{A_1' \cdots A_8'} \, F^{A_6' A_7' A_8'}\,.
\end{equation}
In fact, this is one of the two classes of the equation that one derives from the non-relativistic limit of the self-duality condition $\hat{F}^{(5)} = \star \hat{F}^{(5)}$, the other one being
\begin{equation}\label{eq:longselfdual}
    F_{A A_1' \cdots A_4'} = \frac{1}{4!} \, \epsilon_{AB}^{\phantom{A}} \, \epsilon^{\phantom{A}}_{A_1' \cdots A_4' B_1' \cdots B_4'} \, F^{B B_1' \cdots B_4'}\,.
\end{equation}
Thus, the original self-duality condition, which was imposed by hand, does not arise as an equation of motion. However, in the non-relativistic string limit, it splits into two equations, one of which is the constraint being imposed by the Lagrange multiplier. 

\paragraph{Finite Terms.} Now, we collect the $O( \omega^0 )$ terms in the action \eqref{eq:somegaexp}:
\begin{align} \label{eq:(0)S}
    \stackrel{(0)}{S} = \frac{1}{16 \pi G^{}_\text{N}} & \int \! d^{10} x \, E \, \biggl[ R + \tau^{}_{A'A}{}^{A} \Bigl( 2 \, \tau^{A'B}{}_{B} + \partial^{A'} \Phi \Bigr)
    - \tfrac{3}{8} \, \partial_{A'} \Phi \, \partial^{A'} \Phi - \tfrac{1}{2 \cdot 3!} \, e^{- \Phi} H^{}_{\!A'B'C'} H^{A'B'C'} \notag \\[6pt]
    & + e^{- \Phi / 2} \, \epsilon^{}_{AB} \, \tau^{}_{A'B'}{}^{A} \, H^{A'B'B} - \tfrac{1}{2} \, e^{2 \Phi} \, F^{}_A \, F^A - \tfrac{1}{2} \, e^{3 \Phi / 2} \, F^{A'} F^{}_{A'AB} \, \epsilon^{AB} \notag \\[6pt]
    &  
    - \tfrac{1}{4} \, e^{\Phi} \, F^{}_{A'B'A} \, F^{A'B'A} - \tfrac{1}{4!} \, e^{\Phi/2} \, F^{}_{A'B'C'} \, F^{A'B'C'AB} \, \epsilon^{}_{AB} - \tfrac{1}{4 \cdot 4!} \, F^{}_{A_1' \cdots A_4' A} \, F^{A_1' \cdots A_4'A} \biggr] \notag \\[2pt]
    - \frac{1}{32 \pi G^{}_\text{N}} \, & \int \CC^{(4)} \wedge H^{(3)} \wedge F^{(3)}\,,
\end{align}
where $R$ is the $O( \omega^{1/2} )$ term in $\hat{R}$. In appendix \ref{app:eiibsa}, we show more details on deriving Eq.~\eqref{eq:(0)S} from the Lorentzian action \eqref{eq:relaction}. In appendix \ref{sec:Ricci}, we give the explicit expression and some properties of the scalar curvature $R$\,.
Combining the $O( \omega^0)$ terms in Eq.~\eqref{eq:HS} and Eq.~\eqref{eq:(0)S},
we find that the final bosonic part of the NL IIB supergravity action is given by
\be \label{eq:NLIIB}
    S = \stackrel{(0)}{S} + \frac{1}{16 \pi G^{}_\text{N}} \int \CA^{(5)} \wedge \star \bigl( \Omega^{(3)} \wedge \ell^{(2)} \bigr)\,,
\ee
where the first term is given by Eq.~\eqref{eq:(0)S} and $\Omega^{(3)}$ in Eq.~\eqref{eq:NLIIB} is defined in Eq.~\eqref{eq:Omega3}. 

Unlike the Lorentzian supergravity action \eqref{eq:relaction}, the NL action \eqref{eq:NLIIB} that we obtained via performing the non-relativistic string limit appears to be rather complicated. In particular, the SL($2\,, \mathbb{R}$) symmetry is not as manifest as in the original Lorentzian action \eqref{eq:relaction} anymore. In section~\ref{sec:prsl2rit}, we will discover a more natural way of parametrizing the differential forms in NL IIB supergravity that is distinct from the Lorentzian theory. In this new basis of differential forms, the SL($2\,,\mathbb{R}$) symmetry also becomes manifest in the NL corner. To start with, we first discuss the symmetries of the action \eqref{eq:NLIIB} in the next subsection, which will motivate the formalism in section~\ref{sec:prsl2rit} that we will refer to as a polynomial realization of SL($2\,,\mathbb{R}$).

\subsection{Symmetries in Non-Lorentzian IIB Supergravity}\label{sec:symmetries}

In this section, we investigate the symmetries of the NL IIB action \eqref{eq:NLIIB}. We will first discuss the counterparts of the symmetries that already exist in Lorentzian IIB supergravity, including the higher-form gauge symmetries, spacetime diffeomorphism, rotation, and boost symmetries, and, finally, the global SL($2\,,\mathbb{R}$) symmetries. In addition, NL IIB supergravity is also invariant under a local dilatation symmetry that requires special attention.    

\paragraph{Higher-Form Gauge Symmetries.} We first deal with the higher-form gauge symmetries, which take on the same form as in Lorentzian IIB supergravity and act on the $B$-field and Ramond-Ramond potentials as
\be
    \delta^{}_\xi B^{(2)} = d\xi^{(1)}\,,
        \qquad%
    \delta^{}_\zeta C^{(q)} = d\zeta^{(q-1)} + dB^{(2)}\wedge \zeta^{(q-3)}\,.
\ee
It is understood that $\zeta^{(q)}$ vanishes when $q<0$\,.

\paragraph{Spacetime Gauge Symmetries.} Next, we consider the geometric data in NL IIB supergravity, which are encoded by the NS-NS fields, including the longitudinal Vielbein $\tau^{}_\mu{}^A$\,, transverse Vielbein $E^{}_\mu{}^{A'}$, Kalb-Ramond field $B^{}_{\mu\nu}$\,, and dilaton $\Phi$\,. These fields form the so-called \emph{torsional string Newton-Cartan geometry} \cite{Bergshoeff:2021bmc, Yan:2021lbe, Bidussi:2021ujm}, which is characterized not only by curvature but also intrinsic torsion associated with $\tau^{}_\mu{}^A$ \cite{Bergshoeff:2022fzb}.\footnote{ The underlying spacetime symmetry algebra is known to be the fundamental string Galilei algebra \cite{Bidussi:2021ujm}, the gauging of which gives rise to all the NS-NS fields in non-Lorentzian IIB theory.} However, when the spacetime symmetry transformations of the IIB data are concerned, it is sufficient for us to focus on the string Galilei algebra, which consists of generators associated with longitudinal and transverse translations, longitudinal SO($1\,,1$) Lorentz boost, transverse SO(8) rotations, and string Galilei boosts between the longitudinal and transverse sectors. We collect the spacetime symmetries below:
\begin{enumerate}[(1)]
    
    \item
    
    \emph{Diffeomorphisms}, which are manifestly preserved by appropriately contracting the curved spacetime Greek indices such as $\mu$\,.
    
    \item

    \emph{Local $\text{\emph{SO}}(1\,,1) \times \text{\emph{SO}}(8)$ rotations}, which act infinitesimally on the Vielbeine fields as 
    \be
        \delta^{}_\text{R}\tau_\mu{}^A = \lambda^A{}_B \, \tau_\mu{}^B\,,
            \qquad%
        \delta^{}_\text{R} E_\mu{}^{A'} = \lambda^{A'}{}_{B'} \, E_\mu{}^{B'}\,.
    \ee
    This symmetry is manifestly preserved by appropriately contracting the frame Latin indices such as $A$ and $A'$ in the longitudinal and transverse sector, respectively.

    \item

    \emph{Local string Galilei boosts}, which act on the NS-NS fields as 
    \begin{subequations} \label{eq:sgbteb}
    \begin{align} \label{eq:sgbteb1}
        \delta^{}_\text{G} \tau^{}_\mu{}^A & = 0\,,
            &
        \delta^{}_\text{G} E^{}_\mu{}^{A'} & = -\lambda_A{}^{A'} \, \tau^{}_\mu{}^A\,, 
            &
        \delta^{}_\text{G} B^{(2)} & = \epsilon_{AB} \, \lambda^{B}{}_{A'} \, e^{\Phi/2} \, \tau^A \wedge E^{A'}.
    \end{align}
    From the first two boost transformations in \eqref{eq:sgbteb1}, we find that the inverse Vielbeine fields transform as
    $\delta^{}_\text{G} \tau^\mu{}^{}_A = \lambda_A{}^{A'} \, E^\mu{}^{}_{A'}$ and $\delta^{}_\text{G}\, E^\mu{}^{}_{A'} = 0$\,.
    Note that the string Galilei boost also acts non-trivially on the RR fields $C^{(q)}$, with
    \be
        \delta^{}_\text{G} C^{(q)} = - \epsilon_{AB} \, \lambda^{B}{}_{A'} \, e^{\Phi/2} \, \tau^A \wedge E^{A'} \wedge C^{(q-2)}\,.    
    \ee
    It is understood that $\delta^{}_\text{G} C^{(0)} = 0$\,. Finally, performing the $\omega \rightarrow \infty$ limit of the string Galilei boost transformation of Eq.~\eqref{eq:A5solution} with $\lambda_{A}{}^{A'} = \omega \, \Lambda_{A}{}^{A'}$, we find that
    the infinitesimal boost transformation of the auxiliary field $\CA^{(5)}$ appearing in Eq.~\eqref{eq:NLIIB} is
    \bea
        \delta_G\, \CA^{(5)} = 2 \,\epsilon_{AB}\,\lambda^A{}_{A'}\,E^{A'}\wedge \tau^B\, \wedge \,\Omega^{(3)}\,,
    \eea 
    \end{subequations}
    where $\Omega^{(3)}$ is defined in Eq.~\eqref{eq:Omega3} and satisfies $\Omega^{(3)} \wedge \ell^{(2)} = 0$ as in Eq.~\eqref{eq:o3wl2}. The above boost transformations are also valid for finite $\lambda^{}_A{}^{A'}$\,, as long as ``$\delta$" is interpreted as the finite difference between the transformed and original fields.

    Note that the NL self-duality conditions transform under Galilei boosts as follows:
\be
    \delta^{}_\text{G} F^{}_{A'_1\cdots A'_5} = 0\,,
        \qquad%
    \delta^{}_\text{G} F^{}_{AA'_1\cdots A'_4} = \lambda_A{}{}^{A'_5} \, F^{}_{A'_1\cdots A'_5} -4\, e^{\phi/2}\, \epsilon_{AB}\,\lambda^{B}{}_{[A'_1}\, F_{A'_2\,A'_3\, A'_4]}\,,
\ee
\emph{i.e.}, Eq. \eqref{eq:transselfdual} is invariant, whereas Eq.~\eqref{eq:longselfdual} transforms to Eq.~\eqref{eq:transselfdual}.

\end{enumerate}

\paragraph{Global SL($2\,,\mathbb{R}$) Symmetry.} Let us now turn to the SL($2\,,\mathbb{R}$) symmetry of the NL IIB action \eqref{eq:NLIIB}.  
As shown in \cite{Bergshoeff:2022iss}, the set of SL($2\,,\mathbb{R}$) tranformations in NL IIB theory contains a branching factor that depends on the sign of $\gamma\,C^{(0)} + \delta$ and take the following form:\,\footnote{In \cite{Bergshoeff:2022iss}, we worked with string-frame fields, which explains the different powers of the dilaton in the expressions given here. Moreover, we worked with SL($2\,,\mathbb{Z}$) instead of SL($2\,,\mathbb{R}$) there, which does not change the form of the transformation rules. Finally, the SL($2\,,\mathbb{Z}$) transformations are written for $C^{(4)}$ in \cite{Bergshoeff:2022iss}. Here, we will instead write the SL($2\,,\mathbb{R}$) transformation for $\CC^{(4)}$ in Eq.~\eqref{eq:defCC4}.}
\begin{subequations} \label{eq:globalsl2r}
    \begin{align}
    \tau_\mu{}^{A} &\rightarrow \tau_\mu{}^{A}\,,\hskip 3.5truecm E_\mu{}^{A'} \rightarrow E_\mu{}^{A'}\,,\\[4pt]
    C{}^{(0)} & \rightarrow \frac{\alpha \, C^{(0)} + \beta}{\gamma \, C^{(0)} + \delta}\,,
        \hskip 2.truecm\,\,\,\,\,%
    \Phi \rightarrow \Phi + 2 \, \ln |\gamma \, C^{(0)} + \delta|\,, \label{eq:c0phisl2r} \\[4pt]
    \begin{pmatrix}
        B^{(2)} \\[2pt]
        C^{(2)}
    \end{pmatrix}
    &\rightarrow
    \sgn\big(\gamma\,C^{(0)}+\delta\big)\Bigg[\bigl( \Lambda^{-1} \bigr)^\intercal \, \begin{pmatrix}
        B^{(2)} \\[2pt]
        C^{(2)}
    \end{pmatrix} + \CW\Bigg]\,,\label{eq:bctf}\\[4pt]
    \CC^{(4)} &\rightarrow \CC^{(4)} - \frac{\kappa}{2} \,\Bigl[e^{-\Phi/2}B^{(2)}  - \frac{\kappa}{2} \,  e^{\Phi/2} \bigl( C^{(2)} + C^{(0)}\,B^{(2)} \bigr) \Bigr]\wedge \ell^{(2)}\,,\label{eq:c4trnsf}
    \end{align}
\end{subequations}
where $\Lambda$ is the matrix of \SLR\ parameters $\alpha,\,\beta,\,\gamma,\,\delta$ as given in \eqref{eq:SLParameter}, and 
\begin{align}
        &\kappa = \frac{\gamma\,e^{-\Phi}}{\gamma\,C^{(0)}+\delta}\,,
        &&\CW = - \frac{\gamma}{\gamma \, C^{(0)} + \delta} \ls \bigl( \Lambda^{-1} \bigr)^\intercal + \frac{\mathbb{1}}{\gamma \, C^{(0)} + \delta} \rs
    \begin{pmatrix}
        0 \\[2pt]
        \frac{1}{2} \, \ell^{(2)} \, e^{-3\Phi/2}
    \end{pmatrix}.\label{eq:kappaa3}
\end{align}
At this point, $\kappa$ and $\CW$ are merely useful combinations of fields and parameters. We will see in the next section that these combinations actually play a fundamental role when a systematic examination of the realization of \SLR~in the NL supergravity is concerned. Moreover, we will also introduce a new basis of fields in terms of which the SL($2\,,\mathbb{R}$) transformation rules in Eq.~\eqref{eq:globalsl2r} significantly simplify.  Note that we have used $\CC^{(4)}$ rather than $C^{(4)}$. The transformation rule for $C^{(4)}$ is more complicated and can be found by using the definition of $\CC^{(4)}$ in Eq.~\eqref{eq:defCC4} (see also \cite{Bergshoeff:2022iss}). The SL($2\,,\mathbb{R}$) transformation of the Lagrange multiplier $\CA^{(5)}$\, can be found by performing the SL($2\,,\mathbb{R}$) transformation of Eq.~\eqref{eq:A5solution}, followed by taking the $\omega \rightarrow \infty$ limit. This gives 
\begin{align}\label{eq:A5slr}
    \mathcal{A}{}^{(5)}
        \rightarrow%
    \mathcal{A}{}^{(5)} - \Bigl( \kappa \, e^{-\frac{\Phi}{2}} \, H^{(3)} - \tfrac{1}{2} \, \kappa^2 \, e^{\frac{\Phi}{2}} \, F^{(3)} \Bigr) \wedge\ell^{(2)}\,.
\end{align}

Note that the above expressions of SL($2\,,\mathbb{R}$) transformations are only valid if $\gamma \, C^{(0)} + \delta \neq 0$\,. In the case where $\gamma \, C^{(0)} + \delta = 0$\,, NL IIB supergravity is mapped to the ``one-brane limit" of Lorentzian IIB supergravity. In the associated type IIB superstring theory, this one-brane limit is defined by fine-tuning the RR two-form to cancel the D1-brane tension. This limit should be distinguished from the nonrelativistic string limit, where the Kalb-Ramond two-form is fine-tuned to cancel the fundamental string tension.
Therefore, the SL($2\,,\mathbb{Z}$) transformations satisfying $\gamma \, C^{(0)} + \delta = 0$ map nonrelativistic type IIB superstring theory to the one-brane limit of IIB string theory. The one-brane limit of D1-branes is closely related to Matrix string theory \cite{Banks:1996my, Motl:1997th, Dijkgraaf:1997vv}. See \cite{Ebert:2023hba, uduality} for further details. T-dualities between more general $p$-brane limits associated with various critical RR fields have been discussed in \cite{Gopakumar:2000ep, Danielsson:2000gi} and will be further explored in \cite{uduality, wfmt}, where a duality web unifying a zoo of decoupling limits of string/M-theory, including Matrix (gauge) theories, will be uncovered.

\paragraph{Local Anisotropic Dilatation Symmetry.} NL IIB supergravity also enjoys an emergent local dilatation symmetry parametrized by $\lambda^{}_\text{D}$ that has no counterpart in the parent Lorentzian theory. Such local dilatation acts on an operator $\CO$ as
\be \label{eq:dtrnsf}
    \CO \rightarrow e^{\Delta(\CO) \, \lambda^{}_\text{D}} \, \CO\,,
\ee
where $\Delta(\CO)$ is the dilatation weight associated with the operator $\CO$\,. These weights can be conveniently determined by the exponent in the factor $\omega^\Delta$ in front of each operator in the reparametrization \eqref{eq:NLredefinitions}, with
\begin{subequations} \label{eq:dws}
\begin{align}
    \Delta\big(\tau_\mu{}^A\big) &= - \Delta\big(\tau_A{}^\mu\big) = \frac{3}{4}\,,
        &
    \Delta\big(B^{(2)}\big) &= \Delta\bigl(C^{(q)}\bigr) = 0\,,\\[4pt]
    \Delta\big(E_\mu{}^{A'}\big) &= - \Delta\big(E_{A'}{}^\mu\big) = -\frac{1}{4}\,, 
        &
    \Delta\big(\mathcal A^{(5)}\big) & =0\,, 
        \qquad\quad%
    \Delta\big(e^\Phi\big) = 1\,.
\end{align}
\end{subequations}
The associated fields then transform under the dilatation according to Eq.~\eqref{eq:dtrnsf}. The existence of the dilatation symmetry is a profound property of NL supergravity, as the number of independent fields decreases by one (\emph{e.g.}, one may fix the dilatation gauge by setting $e^\Phi$ to a constant). This introduces an additional Noether identity compared to the Lorentzian case and effectively removes the Poisson equation that encodes the instantaneous Newton-like gravitational interaction. In this sense, the NL IIB supergravity action is a pseudo-action. However, the complete list of target-space equations of motion can still be derived by taking the $\omega \rightarrow \infty$ limit of the Lorentzian equations of motion \cite{Bergshoeff:2019pij, Yan:2021lbe}, or, from first principles via evaluating the beta-functions of the worldsheet theory \cite{Gomis:2019zyu,Gallegos:2019icg,Yan:2019xsf}. We will give more discussions on this in section~\ref{sec:outlook}.

\vspace{3mm}

It is a straightforward but tedious procedure to check that the NL IIB action \eqref{eq:NLIIB} is invariant under the above symmetries. In particular, it is a rather involved exercise to verify that Eq.~\eqref{eq:NLIIB} is invariant under the global SL($2\,,\mathbb{R}$) transformations in Eq.~\eqref{eq:globalsl2r}. This is not very satisfactory, especially because SL($2\,,\mathbb{R}$) is a simple group and there must be a more transparent way to understand its group action. Quite surprisingly, the resolution of this apparent conundrum leads to a simple polynomial realization of SL($2\,,\mathbb{R}$), where all the fundamental fields\,\footnote{Except for the scalars $\Phi$ and $C^{(0)}$\,, which form an SL($2\,,\mathbb{R}$) doublet in Eq.~\eqref{eq:ecp} and transform linearly under the group action.} and their associated field strengths in NL IIB supergravity, which turn out to be in a rather different basis from the ones that we have been working with so far, transform as a polynomial in $\kappa$ as defined in Eq.~\eqref{eq:kappaa3}. We will devote the rest of the paper to developing this highly unconventional SL($2\,,\mathbb{R}$) realization.

Another bona fide surprise comes from the fact that the bosonic sector of NL IIB supergravity can be uniquely determined using the aforementioned bosonic symmetries, without referring to the fermionic sector. See more in section~\ref{sec:EFT}.

\subsection{\texorpdfstring{Branched SL($2\,,\mathbb{R}$) and Dilaton Complexification}{Branched SL(2,R) and Dilaton Complexification}} \label{sec:bsl2zdc}

Before we develop the polynomial realization of SL($2\,,\mathbb{R}$) in NL IIB supergravity, we first introduce a trick to simplify the SL($2\,,\mathbb{R}$) transformations of the two-form fields in Eq.~\eqref{eq:bctf} by relocating the branching factor $\sgn \bigl( \gamma \, C^{(0)} + \delta \bigr)$ to the transformation of the dilaton field $\Phi$\,. This simplification will allow us to focus on the polynomial realization of SL($2\,,\mathbb{R}$) in the next section, instead of dragging along various branching factors. 

We start by complexifying the dilaton field $\Phi$ such that its range becomes $\mathbb{R} + 2\pi i \, \mathbb{Z}$\,. Note that this complexification does not affect the real-valuedness of the non-relativistic string coupling $g^{}_s = e^{\langle \Phi \rangle}$\,. We then modify the SL($2\,,\mathbb{R}$) transformation of $\Phi$ in Eq.~\eqref{eq:c0phisl2r} to be
\begin{align} \label{eq:newphisl2z}
\begin{split}
    \Phi & \rightarrow \Phi + 2 \, \ln \bigl( \gamma \, C^{(0)} + \delta \bigr) \\[4pt]
    & = \Phi + 2 \, \ln \bigl| \gamma \, C^{(0)} + \delta \bigr| + \pi i \, \Bigl[ 1 - \sgn \bigl( \gamma \, C^{(0)} + \delta \bigr) \Bigr]\,.
\end{split}
\end{align}
Note that we have chosen a fixed argument of the log such that it only depends on the sign of $\gamma \, C^{(0)} + \delta$\,. 
The dilaton $\Phi$ gains a shift of $2\pi i$ when $\gamma \, C^{(0)} + \delta < 0$\,. This procedure allows us to effectively absorb all the branching dependence into the SL($2\,,\mathbb{R}$) transformation of $\Phi$\,. Requiring the consistency of the SL($2\,,\mathbb{R}$) transformations demands that the group action on the two-forms in Eq.~\eqref{eq:bctf} be modified. In particular, the associativity of the group action requires
the two-form fields transform as
\be \label{eq:sl2zrelbc20}
    \begin{pmatrix}
        B{}^{(2)} \\[2pt]
        C{}^{(2)} 
    \end{pmatrix}
    \rightarrow
    \bigl( \Lambda^{-1} \bigr)^\intercal \, \begin{pmatrix}
        B^{(2)} \\[2pt]
        C^{(2)}
    \end{pmatrix} + \CW\,,
\ee
where the dependence on sgn$(\gamma \, C^{(0)} + \delta)$ is eliminated and $\CW$ was defined in \eqref{eq:kappaa3}.

The importance of the branching behavior of the SL$(2\,,\mathbb{Z})$\,\footnote{Recall that, in string theory, the \SLR\ is restricted to \SLZ\ due to the quantization of the charges.} transformations in non-relativistic IIB superstring theory has been emphasized in \cite{Bergshoeff:2022iss}. In particular, under the original set of SL($2\,,\mathbb{Z}$) transformations in Eq.~\eqref{eq:globalsl2r}, where $\Phi$ is strictly a \emph{real} field,
the $(p\,, q)$-string action in non-relativistic string theory takes the form, 
\begin{align} \label{eq:spq}
\begin{split}
    S_\text{string} = & - \frac{1}{2} \int d^2 \sigma \, e^{\Phi/2} \, \bigl| p - q \, C^{(0)} \bigr| \, \sqrt{-\tau} \, \tau^{\alpha\beta} \, E_{\alpha\beta} \\[4pt]
    & - \int \sgn \bigl(p - q \, C^{(0)} \bigr) \biggl( p \, B^{(2)} + q \, C^{(2)} + \frac{1}{2} \frac{q^2 \, e^{-3\Phi/2} \, \ell^{(2)}}{p - q \, C^{(0)}} \biggr)\,,
\end{split}
\end{align}
where $\sigma^\alpha$\,, $\alpha = 0\,,1$ are the worldsheet coordinates and 
\be
    \tau^{}_{\alpha\beta} = \p^{}_\alpha X^\mu \, \p^{}_\beta X^\nu \, \tau_\mu{}^A \, \tau_\nu{}^B \, \eta^{}_{AB}\,,
        \qquad%
    E^{}_{\alpha\beta} = \p^{}_\alpha X^\mu \, \p^{}_\beta X^\nu \, E^{}_{\mu}{}^{A'} \, E^{}_\nu{}^{B'} \, \delta^{}_{A'B'} 
\ee
are pull-backs from the target-space to the worldsheet. Moreover, $\tau = \det \tau_{\alpha\beta}$ and $\tau^{\alpha\beta}$ is the inverse of $\tau^{}_{\alpha\beta}$\,. We have set the string tension to one. Note that the $(p\,,q)$-string action contains two different branches: It takes different forms depending on the sign of $p - q \, C^{(0)}$\,. Both of these branches are required in order for the full SL($2\,,\mathbb{Z}$) symmetry to be realized. In contrast, the NL IIB supergravity action \eqref{eq:NLIIB} has a single action that is independent of the sign of $p-q\,C^{(0)}$, just like the D3-brane case in \cite{Bergshoeff:2022iss}. This is expected: The supergravity theory should not be sensitive to which $(p\,,q)$-string is coupled to it. 

Intriguingly, using the trick of complexifying the dilaton field and the new set of SL($2\,,\mathbb{Z}$) transformations, where we have Eqs.~\eqref{eq:newphisl2z} and \eqref{eq:sl2zrelbc20} replacing the corresponding ones in Eq.~\eqref{eq:globalsl2r}, the branching factor in the non-relativsitic $(p\,,q)$-string action is also removed. Now, by imposing invariance under the new set of SL($2\,,\mathbb{Z}$) transformations with a complexified $\Phi$\,, Eq.~\eqref{eq:spq} is replaced with
\be\label{eq:complstring}
    S_\text{string} \! = \! - \frac{1}{2} \int \! d^2 \sigma \, e^{\Phi/2} \bigl( p - q \, C^{(0)} \bigr) \, \sqrt{-\tau} \, \tau^{\alpha\beta} \, E_{\alpha\beta} - \int \biggl[ p \, B^{(2)} + q \, C^{(2)} + \frac{q^2 \, e^{-3\Phi/2} \, \ell^{(2)}}{2\,\bigl(p - q \, C^{(0)}\bigr)} \biggr]\,,
\ee
while the NL versions of both the D3-brane and IIB supergravity action remain unbranched. In this way, all the branching behavior of non-relativistic IIB superstring theory is now located in the branching structure of SL$(2\,,\mathbb{Z})$ transformation \eqref{eq:newphisl2z} of the dilaton field $\Phi$\,.\footnote{Note that only the sign of $e^{\Phi/2}$ (and not $e^\Phi$, which is associated to the string coupling) depends on the choice of branch.} We will stick to this new parametrization of the SL($2\,,\mathbb{R}$) transformations throughout the rest of the paper. Note, however, that the original formulation \eqref{eq:spq} is arguably more physical since it makes manifest the underlying branching between strings satisfying $p-q\,C^{(0)}>0$ and their anti-strings satisfying $p-q\,C^{(0)}<0$\,.

\section{\texorpdfstring{Polynomial Realization of SL$(2\,,\mathbb{R})$ and Invariant Theory}{Polynomial Realization of SL(2,R) and Invariant Theory}} \label{sec:prsl2rit}

In section \ref{sec:symmetries}, we derived the NL IIB supergravity action by performing the non-relativistic string limit of its Lorentzian counterpart and classified its underlying symmetries, where the non-trivial ones include the dilatation, string Galilei boost, and SL($2\,,\mathbb{R}$) symmetries. However, the SL($2\,,\,\mathbb{R}$) invariance of the IIB action \eqref{eq:NLIIB} is far from manifest. This is partly due to the fact that the SL($2\,,\mathbb{R}$) transformations of the higher-form gauge fields take a rather complicated form. For example, the Kalb-Ramond field $\hat{B}^{(2)}$ and Ramond-Ramond two-form field $\hat{C}^{(2)}$ form an SL($2\,,\mathbb{R}$) doublet
\be \label{eq:sbc}
    \hat{\Sigma} = 
    \begin{pmatrix}
        \hat{B}^{(2)} \\[2pt]
        \hat{C}^{(2)}
    \end{pmatrix}
\ee
in Lorentzian IIB theory and transform linearly as 
\be \label{eq:slits}
    \hat{\Sigma} \rightarrow \bigl( \Lambda^{-1} \bigr)^\intercal \, \hat{\Sigma}\,,
\ee
where $\Lambda$ is the SL($2\,,\mathbb{R}$) matrix defined in Eq.~\eqref{eq:SLParameter}.
However, after reparametrizing as in Eq.~\eqref{eq:NLredefinitions} and then taking the $\omega \rightarrow \infty$ limit, we find that Eq.~\eqref{eq:slits} induces the much more complicated transformation rules in Eq.~\eqref{eq:bctf} for the two-form fields $B^{(2)}$ and $C^{(2)}$ in NL IIB theory. In other words, these two-form fields in the IIB theory do \emph{not} form any SL($2\,,\mathbb{R}$) doublet anymore. 
This is not surprising: The transformation of the $O( \omega^2 )$ terms in $\hat{B}^{(2)}$ and $\hCtw$ themselves have a subleading $O( \omega^0 )$ piece, which leads to the complicated non-linear and inhomogenous second term in Eq.~\eqref{eq:bctf}. This is a generic property originating from the interplay between the Lorentzian SL($2\,,\mathbb{R}$) transformations and the $\omega \rightarrow \infty$ limit. In the following subsection, we will formulate this observation in a more general way, which will prove to be very useful for later use.

Intriguingly, the somewhat complicated SL($2\,,\mathbb{R}$) transformations in Eq.~\eqref{eq:bctf} simplify drastically upon the changes of variables,
\begin{subequations}\label{eq:CCbasis}
\begin{align}
    \CB^{(2)} &= e^{-\Phi/2} B^{(2)}\,,
    &\CC^{(2)} &= e^{\Phi/2} \lr C^{(2)} + C^{(0)} \, B^{(2)}\rr\,.
\end{align}
Note that the forms of the new definitions $\CB^{(2)}$ and $\CC^{(2)}$ already appeared in the SL($2\,,\mathbb{R}$) transformation \eqref{eq:c4trnsf} of $\CC^{(4)}$\,. Moreover, following \cite{Bergshoeff:2022iss}, it is useful to define\,\footnote{In~\cite{Ebert:2023hba}, we will see that the quantity \eqref{eq:ecp} is a Vielbein field on the anisotropic torus over which non-relativistic M-theory is compactified.} 
\be \label{eq:ecp}
    \begin{pmatrix}
        \text{e}^1 \\[4pt]
        \text{e}^2
     \end{pmatrix}
     = e^{\Phi/2}\,
    \begin{pmatrix}
        C^{(0)}\\[4pt] 
        1
    \end{pmatrix}\,.
\ee
\end{subequations}
In this new basis, and in terms of the complexified dilaton $\Phi$ introduced in section~\ref{sec:bsl2zdc}, we find that 
the SL($2\,,\mathbb{R}$) transformations in Eq.~\eqref{eq:globalsl2r} now become
\begin{subequations} \label{eq:cbcct}
\begin{align}  
    \begin{pmatrix}
        \text{e}^1 \\[4pt]
        \text{e}^2
     \end{pmatrix}
        & \rightarrow \Lambda \begin{pmatrix}
        \text{e}^1 \\[4pt]
        \text{e}^2
     \end{pmatrix}\,, \label{eq:e12slz}
\end{align}
and
\begin{align}
    \CB^{(2)} &\rightarrow \CB^{(2)} - \kappa \, \CC^{(2)} + \tfrac{1}{2} \, \kappa^2 \, \ell^{(2)}\,, \\[4pt]
    \CC^{(2)} &\rightarrow \CC^{(2)} - \kappa \, \ell^{(2)}\,, \\[4pt]
    \CC^{(4)} &\rightarrow \CC^{(4)} - \tfrac{1}{2} \, \kappa \,\CB^{(2)} \wedge \ell^{(2)} + \tfrac{1}{4} \, \kappa^2 \, \CC^{(2)}\wedge \ell^{(2)}\,.
\end{align}
\end{subequations}
We have replaced the $\Phi$ transformation in Eq.~\eqref{eq:c0phisl2r} with Eq.~\eqref{eq:newphisl2z} and replaced Eq.~\eqref{eq:bctf} with Eq.~\eqref{eq:sl2zrelbc20}.
Here,
\be \label{eq:kappadef}
    \kappa = \frac{\gamma \, e^{-\Phi}}{\gamma \, C^{(0)} + \delta}\,.
\ee
has previously appeared in Eq.~\eqref{eq:kappaa3}. Together with the transformation rule of the Lagrange multiplier $\CA^{(5)}$ in \eqref{eq:A5slr}, we find that the full SL($2\,,\mathbb{R}$) transformations are now expressed in terms of polynomials in $\kappa$ as in Eq.~\eqref{eq:cbcct}.
As we will demonstrate later in this section, similar reparametrizations exist for all the associated field strengths, such that the reparametrized field strengths also transform under the SL($2\,,\mathbb{R}$) action as a polynomial in $\kappa$\,. We reemphasize that the formalism we are developing here is only valid for $\gamma \, C^{(0)} + \delta \neq 0$\,. See the comments below Eq.~\eqref{eq:A5slr}.

It is tempting to ask whether there is any profound mathematical structure underlying the above observation. We will answer this question through this section and discover a polynomial realization of the SL($2\,,\mathbb{R}$) transformations. We will also show that this novel realization of the SL($2\,,\mathbb{R}$) action is closely related to a natural generalization of the classical invariant theory of polynomial equations and binary forms (homogeneous polynomials in two variables) \cite{olver1999classical}. Later in section~\ref{sec:anl2bs}, we will show how the SL($2\,,\mathbb{R}$) invariants in NL IIB supergravity can be constructed from two simple binary forms. These results match the expressions from performing the $\omega \rightarrow \infty$ limit in Lorentzian IIB supergravity, but have the benefit of making the SL($2\,,\mathbb{R}$) invariance of the NL supergravity action manifest. 

Finally, we note that the polynomial realization of SL($2\,,\mathbb{R}$) to be developed in this section acquires an elegant geometric interpretation in the context of non-relativistic M-theory that uplifts non-relativistic superstring theory to eleven dimensions. We refer the interested readers to the upcoming paper \cite{Ebert:2023hba}.  

\subsection{Global Symmetries in Non-Lorentzian Theories} \label{eq:gsnlt}

We start by extracting the general concept behind how the more complicated SL($2\,,\mathbb{R}$) transformations such as Eq.~\eqref{eq:bctf} in NL IIB theory arise from the $\omega \rightarrow \infty$ limit of the linear transformations in Lorentzian IIB theory. This abstraction will reveal the underlying mathematical structure of the SL($2\,,\mathbb{R}$) invariants in NL IIB supergravity. We start with a heuristic argument to motivate our central expression \eqref{eq:condsK0}, which will play an essential role in our later construction of the polynomial realization of SL($2\,,\mathbb{R}$) in section~\ref{sec:prslzit}. 

Consider a Lorentzian system and denote the space of all its operators by $\hat{\mathscr{O}}$\,. We assume that all operators in  $\hat{\mathscr{O}}$ are covariant under Lorentz transformations. Furthermore, we require that this physical system be invariant under the action of a global group $G$\,, which acts linearly on a certain operator $\hat{\CO} \in \hat{\mathscr{O}}$ as
\be \label{eq:otrnsf}
    \hat{\CO} \rightarrow g \cdot \hat{\CO}\,,
        \qquad%
    g \in G\,.
\ee
In general, the group action ``$\cdot$" can have nonlinear dependence on other operators in $\hat{\mathscr{O}}$\,. For simplicity, we will restrict to the subspace of $\hat{\mathscr{O}}$ on which the above group action is linear in all of the operators. In the example we considered at the beginning of this section, we have $G = \text{SL}(2\,,\mathbb{R})$ and Eq.~\eqref{eq:otrnsf} becomes $\hat\CO\to(\Lambda^{-1})^\intercal \, \hat{\CO}$ as in Eq.~\eqref{eq:slits}, with $\hat{\CO}$ being identified with the SL($2\,,\mathbb{R}$) doublet $\hat{\Sigma}$ in Eq.~\eqref{eq:sbc}. 

Next, we consider a reparametrization of the fundamental fields that form all the operators in $\hat{\mathscr{O}}$\,, which takes the form of\,\footnote{More general Ans\"{a}tze of Eq.~\eqref{eq:om2o} may be considered (\emph{e.g.}, a reparametrization of the operators containing higher powers of $\omega$). However, it is sufficient to stick to Eq.~\eqref{eq:om2o} for our purpose. Moreover, one may even consider an expansion of all the operators with respect to a large $\omega$ such that subleading orders in $\omega$ are also included in Eq.~\eqref{eq:om2o} (see, \emph{e.g.}, \cite{Hartong:2021ekg, Hartong:2022dsx}). 
Nevertheless, since we will be taking an $\omega \rightarrow \infty$ limit, it is sufficient that we redefine the finite piece in the expansion to absorb all the subleading-order terms.} 
\be \label{eq:om2o}
    \hat{\CO} = \omega^2 \, \CO_0 + \CO\,.
\ee
Note that the reparametrizations in Eq.~\eqref{eq:NLredefinitions} provide an explicit realization of Eq.~\eqref{eq:om2o}.
We require such a reparametrization of the operator space to be invertible. It is important that the reparametrizations like Eq.~\eqref{eq:om2o} contain a term divergent at infinite $\omega$\,. This is key to the breaking of Lorentzian symmetries in the $\omega \rightarrow \infty$ limit, under the condition that the Vielbeine fields encoding the spacetime geometry are also rescaled by $\omega$ anisotropically in space and time. After the $\omega \rightarrow \infty$ limit is performed, we denote the resulting operator space by $\mathscr{O}$\,. Note that the unhatted operators $\CO_0$ and $\CO$ are both elements in $\mathscr{O}$\,.  

Expanding the group action \eqref{eq:otrnsf} with respect to a large $\omega$ using the reparametrization \eqref{eq:om2o}, we find
\be \label{eq:co0cotrnsf}
    \CO_0 \rightarrow g \cdot \CO_0 + O(\omega^{-2})\,,
        \qquad%
    \CO \rightarrow g \cdot \CO + \CK(g\,, \mathscr{O}) + O (\omega^{-2})\,.
\ee
The shift $\CK(g\,, \mathscr{O})$ in Eq.~\eqref{eq:co0cotrnsf} is a function of both group parameters and operators in the NL theory, and it arises due to the fact that the subleading group transformation in $\mathcal{O}_0$ is generically nonzero and that the group action itself may depend on $\omega$\,. In the $\omega \rightarrow \infty$ limit, we find
\be
    \CO_0 \rightarrow g  \cdot \CO^{}_0\,,
        \qquad%
    \CO \rightarrow g  \cdot \CO + \CK(g\,, \mathscr{O})\,.
\ee
In the above heuristic way, which is already sufficient for the purpose of this paper, we conclude that the group action of interest on the operators in the resulting NL system take the following form:
\be \label{eq:gcogdok}
    g \circ \CO = g  \cdot \CO + \CK(g\,, \mathscr{O})\,,
\ee
where both $g \circ \CO$ and $g \cdot \CO$ are group actions and satisfy the consistency conditions
\begin{subequations} \label{eq:gaaxiom0}
\begin{align}
	\text{Identity}: && \mathbb{1} \circ \CO & = \CO\,, 
		& %
		\mathbb{1} \cdot \CO & = \CO\,, \label{eq:1psi} \\[4pt]
	\text{Compatibility}: && g' \circ \bigl(g \circ \CO\bigr) & = \bigl( g' g \bigr) \circ \CO\,,
		& %
		g' \cdot \bigl(g \cdot \CO\bigr) & = \bigl( g' g \bigr) \cdot \CO\,,
\end{align}
\end{subequations}
with $\mathbb{1}$ the identity in $G$\,. The above conditions imply that $\CK$ satisfies
\be \label{eq:condsK0}
	\CK\bigl(\mathbb{1}\,, \mathscr{O}\bigr) = 0\,,
		\qquad%
	\CK\bigl(g' \, g\,, \mathscr{O}\bigr) = \CK\bigl(g', \, g \circ \mathscr{O}\bigr) + g' \cdot \CK \bigl(g\,, \mathscr{O}\bigr)\,.
\ee
These conditions on $\CK$ will turn out to be crucial to formulating our polynomial realization of SL($2\,,\mathbb{R}$). 

As an example, we rewrite the SL($2\,,\mathbb{R}$) transformations in Eq.~\eqref{eq:bctf} in terms of the two group actions as in Eq.~\eqref{eq:gcogdok}. In this case, 
\be \label{eq:cogoc}
    \CO = \begin{pmatrix}
        B^{(2)} \\[2pt]
        C^{(2)}
    \end{pmatrix}\,,
        \qquad%
    g \cdot \CO = \bigl( \Lambda^{-1} \bigr)^\intercal \, \begin{pmatrix}
        B^{(2)} \\[2pt]
        C^{(2)}
    \end{pmatrix}\,,
        \qquad%
    g \circ \CO = g \cdot \CO + \CW\,,
\ee
with $\CK\bigl(g\,, \mathscr{O}\bigr)=\CW$ and $\CW$ is defined in Eq.~\eqref{eq:kappaa3}.
Note that $\CO$ is an SL($2\,,\mathbb{R}$) doublet with respect to the group action $g \cdot \CO$ and $g' \cdot \CW = \bigl( \Lambda'{}^{-1} \bigr)^\intercal \, \CW$\,.\,\footnote{Note that $\CW$ contains $C^{(0)}$ and $\Phi$\,, for which the group action $\circ$ is strictly speaking not defined. But it is induced by the linear SL($2\,,\mathbb{R}$) group action on $\bigl(\text{e}^{1}, \text{e}^{2} \bigr)^\intercal$ in Eq.~\eqref{eq:e12slz}.}

\subsection{\texorpdfstring{A Polynomial Realization of SL(2,\,$\mathbb{R}$)}{A Polynomial Realization of SL(2,R)}} \label{sec:prslzit}

Eventually, we will consider the application of the formalism motivated in section~\ref{eq:gsnlt} to the SL($2\,,\mathbb{R}$) duality in NL IIB supergravity. For this purpose, we will show later in section~\ref{sec:anl2bs} that it essentially suffices to consider operators that are SL($2\,,\mathbb{R}$) singlets with respect to the group action $g \cdot \CO$\,, except for the transformations of $C^{(0)}$ and $\Phi$ (see Eqs.~\eqref{eq:e12slz} and \eqref{eq:ecp}).\,\footnote{Even though the two-form fields in Eq.~\eqref{eq:cogoc} transform as a doublet with respect to the group action $g \cdot \CO$\,, the new variables $\CB^{(2)}$ and $\CC^{(2)}$ introduced in Eq.~\eqref{eq:CCbasis} are singlets with respect to this group action.} We thus simplify Eq.~\eqref{eq:gcogdok} to be
\be \label{eq:groupaction}
    g \circ \CO = \CO + \CK(g\,, \mathscr{O})\,,
\ee
and simplify the associated consistency conditions \eqref{eq:condsK0} to be
\be \label{eq:consconds}
    \CK\bigl(\mathbb{1}\,, \mathscr{O}\bigr) = 0\,,
		\qquad%
	\CK\bigl(g' \, g\,, \mathscr{O}\bigr) = \CK\bigl(g', \, g \circ \mathscr{O}\bigr) +  \CK \bigl(g\,, \mathscr{O}\bigr)\,. 
\ee
In order to develop a non-trivial realization of SL($2\,,\mathbb{R}$) in this form, it is key to construct a function $\CK$ satisfying Eq.~\eqref{eq:consconds}. 

The quantity $\kappa$ in Eq.~\eqref{eq:kappadef} forms a desired singlet under the ``$\cdot$" group action and satisfies the consistency conditions in Eq.~\eqref{eq:consconds}:\,\footnote{There is a slightly more general expression that also satisfies Eq.~\eqref{eq:consconds},
$$
    \kappa \bigl(g\,, \CO \bigr) = c^{}_1 \, \frac{\beta}{\text{e}^1 \, \bigl(\alpha \, \text{e}^1 + \beta \, \text{e}^2 \bigr)} + c^{}_2 \, \frac{\gamma}{\text{e}^2 \, \bigl( \gamma \, \text{e}^1 + \delta \, \text{e}^2 \bigr)}\,,
$$
where $c^{}_{1}$ and $c^{}_{1}$ are constants. When $c^{}_1 = 0$ and $c^{}_2 = 1$\,, this expression reduces to Eq.~\eqref{eq:kappadef}. It would be interesting to understand whether this is the unique solution to the consistency conditions \eqref{eq:consconds}.}
using Eqs.~\eqref{eq:ecp} and \eqref{eq:e12slz}, we find
\be \label{eq:kappascc}
    \kappa\bigl(\mathbb{1}\,, \mathscr{O}\bigr) = 0\,,
		\qquad%
	\kappa\bigl(g' \, g\,, \mathscr{O}\bigr) = \kappa\bigl(g', \, g \circ \mathscr{O}\bigr) +  \kappa \bigl(g\,, \mathscr{O}\bigr)\,,
\ee
which is precisely Eq.~\eqref{eq:consconds}.
This $\kappa$ plays an essential role in the polynomial realization of SL($2\,,\mathbb{R}$). In fact, we will realize a basis of the operator space such that all the SL($2\,,\mathbb{R}$) transformations are polynomials in $\kappa$\,. 

Note that the self-consistency conditions in Eq.~\eqref{eq:kappascc} are equivalent to the linear transformation \eqref{eq:e12slz} up to a sign. We already showed that Eq.~\eqref{eq:e12slz} implies Eq.~\eqref{eq:kappascc}. Now, we show that Eq.~\eqref{eq:e12slz} can also be recovered from Eq.~\eqref{eq:kappascc}. While the first condition in Eq.~\eqref{eq:kappascc} is automatically satisfied by $\kappa$ in Eq.~\eqref{eq:kappadef}, as $\gamma = 0$ when $\Lambda = \mathbb{1}$\,, plugging Eqs.~\eqref{eq:ecp} and \eqref{eq:kappadef} into the second condition in Eq.~\eqref{eq:kappascc} gives\,
\be
     \Bigl[ \bigl( \alpha \, \text{e}^1 + \beta \, \text{e}^2 \bigr) \bigl( \gamma \, \text{e}^1 + \delta \, \text{e}^2 \bigr) - \bigl( g \circ \text{e}^1 \bigr) \bigl( g \circ \text{e}^2 \bigr) \Bigr] \, \gamma' + \Bigl[ \bigl( \gamma \, \text{e}^1 + \delta \, \text{e}^2 \bigr)^2 - \bigl( g \circ \text{e}^2 \bigr)^2 \Bigr] \, \delta' = 0\,,
\ee
which has to hold for any $\gamma'$ and $\delta'$\,. This implies that both the coefficients in front of $\gamma'$ and $\delta'$ have to vanish individually, which are solved by
\be \label{eq:gce}
    g \circ 
    \begin{pmatrix}
        \text{e}^1 \\[4pt]
        \text{e}^2
     \end{pmatrix}
     =
     \pm \Lambda
     \begin{pmatrix}
        \text{e}^1 \\[4pt]
        \text{e}^2
     \end{pmatrix}.
\ee
The transformations in Eq.~\eqref{eq:gce} with either choice of the sign gives the desired SL($2\,,\mathbb{R}$) rule in Eq.~\eqref{eq:e12slz} (up to a redefinition of the group parameters). This implies that Eq.~\eqref{eq:kappascc} faithfully realizes PSL($2\,,\mathbb{R}$). In our later application to NL IIB supergravity, we will have to include the transformations \eqref{eq:cbcct} of $C^{(0)}$ and $\Phi$\,, which means that the underlying global symmetry is still SL($2\,,\mathbb{R}$)\,.

An $N$-dimensional polynomial realization of the group SL($2\,,\mathbb{R}$) can be constructed by requiring that $\CK$ in Eq.~\eqref{eq:groupaction} be a degree-$N$ polynomial of $\kappa$\,, with
\be \label{eq:polkn}
	\CK = \sum_{m=0}^N \frac{(-\kappa)^m}{m!} \, s^{}_{N-m}(\mathscr{O})\,.
\ee  
Note that $\CK$ has to satisfy the conditions in Eq.~\eqref{eq:consconds}.
Since $\kappa(\mathbb{1}\,, \mathscr{O}) = 0$\,, the first condition from Eq.~\eqref{eq:consconds}, \emph{i.e.}, $\CK(\mathbb{1}\,, \mathscr{O}) = 0$\,, is automatically satisfied. 
The second condition from Eq.~\eqref{eq:consconds} gives
\begin{align} \label{eq:coeffzero}
	0 & = g \circ s^{}_N + \sum_{m=1}^N \frac{1}{m!} \Biggl\{ g \circ s^{}_{N-m} - \sum_{\ell=m}^N 
	\frac{\bigl[-\kappa (g\,, \mathscr{O}) \bigr]^{\ell-m}}{(\ell-m)!} \, s^{}_{N-\ell} \Biggr\} \, \bigl[ -\kappa \bigl( g', \, g \circ \mathscr{O} \bigr) \bigr]^m\,.
\end{align}
Since $g' \in G$ is arbitrary, each coefficient in front of $\kappa^m (g'\,, g \circ \mathscr{O})$\,, $m=0, \cdots, N$ has to vanish identically. This implies $s^{}_N = 0$ because $g \circ s^{}_N = 0$ for all $g \in \text{SL}(2\,,\mathbb{R})$\,, and
\begin{align} \label{eq:gsi0}
	g \circ s_m = \sum_{\ell=0}^{m} 
	\frac{(-\kappa)^{\ell}}{\ell!} \, s_{m-\ell}\,,
        \qquad%
    m = 0\,, \,, \cdots\,, N-1\,,
\end{align}
\emph{i.e.},
\be \label{eq:gsi0matrix}
    g \circ \textbf{S}_N = \textbf{U}_N \, \textbf{S}_N\,,
\ee
where
\be \label{eq:defsu}
    \textbf{S}_N =
    \begin{pmatrix}
        s^{}_0 \\[4pt]
        s^{}_1 \\[4pt]
        s^{}_2 \\[4pt]
        \vdots \\[4pt]
        s^{}_{N-1}
    \end{pmatrix}\,,
        \qquad%
    \textbf{U}_N =
    \begin{pmatrix}
        1 & 0 & 0 &\,\, \cdots &\,\,\, 0\,\, \\[4pt]
        -\kappa & 1 & 0 &\,\, \cdots &\,\,\, 0\,\, \\[4pt]
        \frac{\kappa^2}{2} & -\kappa & 1 & \cdots & 0 \\[4pt]
        \vdots &\,\, 
        \vdots & \vdots &\,\, \ddots &\,\,\, \vdots\,\, \\[4pt]
        \frac{(-\kappa)^{N-1}}{(N-1)!} &\,\, \frac{(-\kappa)^{N-2}}{(N-2)!} &\,\, \frac{(-\kappa)^{N-3}}{(N-3)!} &\,\, \cdots &\,\,\, 1\,\,
    \end{pmatrix}\,.
\ee
The vector $\mathbf{S}_N$ forms an $N$-dimensional realization of SL($2\,,\mathbb{R}$)\,. Since the group transformation of $s^{}_m$ is a degree-$m$ polynomial in $\kappa$\,, we refer to $\mathbf{S}_N$ as a \emph{polynomial realization} of SL($2\,,\mathbb{R}$)\,.

We have already seen an example of the polynomial realization \eqref{eq:gsi0matrix} in Eq.~\eqref{eq:cbcct}. Since $\ell^{(2)}$ is invariant under SL($2\,,\mathbb{R}$), the transformations in Eq.~\eqref{eq:cbcct} can be rewritten as
\be\label{eq:3dsl2fields}
    g \circ
    \begin{pmatrix}
        \ell^{(2)} \\[4pt]
        \CC^{(2)} \\[4pt]
        \CB^{(2)}
    \end{pmatrix}
    =
    \begin{pmatrix}
        1 &\quad 0 &\quad\,\, 0 \\[4pt]
        - \kappa &\quad 1 &\quad\,\, 0 \\[4pt]
        \frac{1}{2} \, \kappa^2 &\quad - \kappa &\quad\,\, 1 
    \end{pmatrix}
    \begin{pmatrix}
        \ell^{(2)} \\[4pt]
        \CC^{(2)} \\[4pt]
        \CB^{(2)}
    \end{pmatrix}\,. 
\ee
Here, $\bigl( \ell^{(2)}\,, \, \CC^{(2)}\,, \, \CB^{(2)} \bigr)^\intercal$ is a three-dimensional polynomial realization of SL($2\,,\mathbb{R}$). For the precise connection to IIB supergravity, we refer to section \ref{sec:anl2bs}. 

Before diving deeper into the mathematical structure of the polynomial realizations, we provide a simple application to the formalism of the $(p\,,q)$-string action \eqref{eq:complstring} in non-relativistic IIB superstring theory. 
A generic ($p\,,q$)-string state is labeled by two relatively co-prime integers $p$ and $q$ that form a doublet under the SL($2\,,\mathbb{Z}$) action, with
\be
    \begin{pmatrix}
        p \\[2pt]
        q
    \end{pmatrix} 
    \rightarrow
    \Lambda \begin{pmatrix}
        p \\[2pt]
        q
    \end{pmatrix}\,. 
\ee
Note that the group parameters in $\Lambda$ are taken to be integers for SL($2\,,\mathbb{Z}$)\,. Together with the doublet $(\text{e}^1\,, \text{e}^2)^\intercal$ defined in Eq.~\eqref{eq:ecp}, we form the SL($2\,,\mathbb{Z}$) invariant 
\be \label{eq:peinv}
    p \, \text{e}^2 - q \, \text{e}^1 = e^{\Phi/2} \bigl( p - q \, C^{(0)} \bigr)\,.
\ee
Furthermore, we define
\be \label{eq:chi}
    \chi = - \frac{q \, e^{-\Phi}}{p - q \, C^{(0)}}\,,
\ee
which transforms as $\chi \rightarrow \chi - \kappa$ under SL($2\,,\mathbb{Z}$)\,. The somewhat exotic quantity $\chi$ that we introduced in \eqref{eq:chi} will find a natural M-theory interpretation in \cite{Ebert:2023hba}: it is associated with the auxiliary vector field used to impose the self-duality condition of the three-form gauge potential in the M5-brane action. Together with the three-dimensional realization in Eq.~\eqref{eq:3dsl2fields}, we form the following SL($2\,,\mathbb{Z}$) invariant:
\be \label{eq:bcc}
    \CB^{(2)} - \chi \, \CC^{(2)} + \tfrac{1}{2} \, \chi^2 \, \ell^{(2)}\,.
\ee
In terms of the SL($2\,,\mathbb{Z}$) invariants in Eqs.~\eqref{eq:peinv} and \eqref{eq:bcc}, we find that the CS term of the ($p\,,q$)-string action \eqref{eq:complstring} in non-relativistic string theory is now recast to be 
\be
    S_\text{string}^\text{CS} = \int \Bigl( p \, \text{e}^2 - q \, \text{e}^1 \Bigr) \Bigl( \CB^{(2)} - \chi \, \CC^{(2)} + \tfrac{1}{2} \, \chi^2 \, \ell^{(2)} \Bigr)\,. 
\ee
where it is understood that the two-forms have been pulled back from the target space to the $(p\,,q)$-string worldsheet. The reformulation of the manifestly SL($2\,,\mathbb{Z}$) invariant D3-brane action in non-relativistic string theory in terms of our new basis, and its origin from non-relativistic M-theory, can be found in \cite{Ebert:2023hba}.

\subsection{Generalized Invariant Theory of Binary Forms} \label{sec:gitbf}

The polynomial realization of SL($2\,,\mathbb{R}$) in section~\ref{sec:prslzit} is intimately related to the mathematics of classical invariant theory that studies the geometric properties of polynomials (see, \emph{e.g.}, \cite{olver1999classical} for a pedagogical introduction). These geometric properties are, by definition, unaffected by certain changes of variables. For example, multiplicities of the roots of a polynomial are geometric but the explicit values of the roots are not. A systematic study of these geometric properties requires finding a set of variables such that the polynomials have a simple, canonical form. This is essentially a problem of equivalence: how do polynomials transform into each other without changing their geometric properties? In order to understand these fundamental questions, it is extremely useful to classify invariants (and covariants) formed by the coefficients in the polynomials: such objects characterize the underlying intrinsic properties of a polynomial.

The simplest example of invariant theory concerns a quadratic binary form that is a homogeneous polynomial $\CP (x\,, y)$ in two variables $x$ and $y$\,, 
\be \label{eq:ihp}
    \CP (x\,, y) = p^{}_0 \, x^2 + 2 \, p^{}_1 \, x \, y + p^{}_2 \, y^2 \,.
\ee
We focus on the case with real coefficients $p^{}_m \in \mathbb{R}$\,. The general transformations preserving the quadratic form are the invertible linear changes
of variables,
\begin{align} \label{eq:gl2r}
    \begin{pmatrix}
        x \\[2pt]
        y
    \end{pmatrix} 
    \rightarrow%
    \begin{pmatrix}
        \alpha &\quad \beta \\[2pt]
        \gamma &\quad \delta
    \end{pmatrix}
    \begin{pmatrix}
        x \\[2pt]
        y
    \end{pmatrix}\,,
        \qquad%
    \alpha \, \delta - \beta \, \gamma \neq 0\,,
\end{align}
which form the group GL$(2\,,\,\mathbb{R})$\,. Further imposing the unimodularity condition $\alpha \, \delta - \beta \, \gamma = 1$\,, this group reduces to SL($2\,,\mathbb{R}$)\,.
Requiring that the quadratic binary \eqref{eq:ihp} be invariant under the SL($2\,,\mathbb{R}$) transformation induces a group action on the coefficients $p^{}_{0}$\,, $p^{}_1$\,, and $p^{}_2$\,, which form a three-dimensional representation of SL($2\,,\mathbb{R}$). Any SL($2\,,\mathbb{R}$) invariant formed by $p^{}_{0}$\,, $p^{}_1$\,, and $p^{}_2$ can be written as a polynomial of a single fundamental invariant, which is the discriminant $\Delta = p_1^2 - p^{}_0 \, p^{}_2$ of the quadratic binary. One major task of invariant theory concerns the classification of the basis of invariants for systems of binary or multi-variable forms at higher degrees. Many fascinating methods for such a classification have been developed in the 19th century, which eventually led to its grand finale marked by Hilbert's proof of the finiteness theorem \cite{hilbert1970ueber}. This theorem shows that any finite system of homogeneous polynomials has a finite basis for its invariants. In the last century, based on the work by Hilbert \cite{hilbert1970vollen}, invariant theory was further advanced by Mumford, which led to geometric invariant theory \cite{mumford1994geometric}. This more advanced subject focuses on group actions on an algebraic variety and involves techniques from algebraic geometry. In contrast to classical invariant theory, geometric invariant theory is capable of studying the relations between invariants without knowing the complete basis.

\subsubsection{\texorpdfstring{Polynomial Realization of SL$(2\,,\mathbb{R})$ Revisited}{Polynomial Realization of SL(2,R) Revisited}}

We now develop a generalization of classical invariant theory and show how the polynomial realization \eqref{eq:gsi0} of SL($2\,,\mathbb{R}$) can be reproduced. We focus on the degree-$(N-1)$ binary form,
\be
    \CP (x\,, y) = \sum_{m=0}^{N-1} 
    \begin{pmatrix}
        N-1 \\[4pt]
        m
    \end{pmatrix}
    p^{}_m \, x^m \, y^{N-m-1}\,.
\ee
We require that $(x\,, y)^\intercal$ be a two-dimensional polynomial realization of SL($2\,,\mathbb{R}$)\,,
\be \label{eq:gxy}
    g \circ 
    \begin{pmatrix}
        x \\[4pt]
        y
    \end{pmatrix}
    =
    \begin{pmatrix}
        1 &\quad 0 \\[2pt]
        -\kappa &\quad 1
    \end{pmatrix}
    \begin{pmatrix}
        x \\[2pt]
        y
    \end{pmatrix}\,,
\ee
where $\kappa$ is given by Eq.~\eqref{eq:kappadef}.
In order to build an $N$-dimensional polynomial realization $(p^{}_0\,, \, \cdots\,, \, p^{}_{N-1})^\intercal$ of SL($2\,,\mathbb{R}$)\,, we demand
\be \label{eq:cprelation}
    \bigl( g \circ \CP \bigr) \bigl(x\,,\, y\bigr) = \CP \bigl( g \circ x\,, \, g \circ y \bigr)\,,
\ee
which implies
\be \label{eq:gpi}
    g \circ p^{}_m = \sum_{\ell=0}^m \begin{pmatrix}
        m \\[2pt]
        \ell
    \end{pmatrix} 
    \, p^{}_\ell \, (-\kappa)^{m-\ell}\,, 
        \qquad%
    m = 0\,, \,\cdots\,,\, N-1\,.
\ee
Here, $g \circ \CP$ means the group action is on the coefficients $p^{}_m$\,. 
Setting 
\be
    p^{}_m = m! \, s^{}_m\,, 
\ee
we find that Eq.~\eqref{eq:gpi} coincides with Eq.~\eqref{eq:gsi0}, \emph{i.e.}, the polynomial realization of SL($2\,,\mathbb{R}$) is recovered. We note a major distinction between this construction and the standard classical invariant theory of binary forms: we are interested in the group action \eqref{eq:gxy} on the variables $x$ and $y$ that depend nonlinearly on the background fields via $\kappa$\,, instead of the linear changes of variables in Eq.~\eqref{eq:gl2r}. However, as far as the classification of invariants is concerned, $\kappa$ in Eq.~\eqref{eq:gxy} can be effectively replaced with an arbitrary real parameter $k \in \mathbb{R}$\,. From this perspective, the group  \eqref{eq:gxy} spanned by
\be
    N =
    \begin{pmatrix}
        1 &\quad 0 \\[4pt]
        -k &\quad 1
    \end{pmatrix}\,,
        \qquad%
    k \in \mathbb{R}
\ee
is the unipotent subgroup of SL($2\,, \mathbb{R}$) in the Iwasawa decomposition \cite{iwasawa1949some}.\,\footnote{We would like thank Niels Obers for pointing out this connection to the Iwasawa decomposition.} As a result, the invariants under the group action \eqref{eq:gxy} in general proliferate compared to the ones under the linear changes of variables \eqref{eq:gl2r}. 
Another distinction is that the variables $x\,, y$ and the coefficients $p^{}_0\,, \, \cdots\,, \, p^{}_{N-1}$ are supposed to be elements of a \emph{field}, which in invariant theory is usually taken to be real or complex numbers. However, for our applications in NL IIB supergravity, these quantities are supposed to be differential forms. 

\subsubsection{Classification of Quadratic Invariants} \label{sec:cqi}

Ultimately, in section~\ref{sec:sl2rinl}, we are interested in constructing the SL($2\,,\mathbb{R}$) invariants in the NL IIB action that are quadratic in spacetime derivatives. In this special case, components in a vector $\mathbf{S}^{}_N$ will be identified with various differential forms representing the field strengths. As far as an $N$-dimensional representation of SL($2\,,\mathbb{R}$) is concerned, this boils down to the classification of expressions quadratic in $s^{}_m$\,, $m=0\,,\, \cdots\,,\,N-1$ that are invariant under the transformation \eqref{eq:gsi0matrix} (or, equivalently, Eq.~\eqref{eq:gpi} in terms of $p^{}_m$). Since these coefficients $s^{}_m$ are differential forms in NL IIB supergravity, we are required to introduce an inner product $\langle s^{}_\ell\,, s^{}_m \rangle$\,. The space of the quadratic invariants is spanned by the basis of SL($2\,,\mathbb{R}$) invariants
\be \label{eq:Im}
	I_r = \sum_{m=0}^{2r} (-1)^m \, \Bigl\langle s^{}_m\,,  s^{}_{2r-m} \Bigr\rangle\,,
		\qquad%
	r = 0\,, 1\,, \, \cdots\,, \lfloor \tfrac{1}{2} \bigl( N-1 \bigr) \rfloor\,,
\ee
where $\lfloor\cdots\rfloor$ is the floor function. Any SL($2\,,\mathbb{R}$) invariant quadratic in $s^{}_m$ can be written as a linear combination of the basis elements in Eq.~\eqref{eq:Im}. The invariance $g \circ I_r = I_r$ can be seen by noting that
\be \label{eq:irsas}
    I^{\phantom{\intercal}}_r = \Bigl\langle \textbf{S}^\intercal_{2r+1} \,, \, \textbf{A}^{\phantom{\intercal}}_r \, \textbf{S}^{\phantom{\intercal}}_{2r+1} \Bigr\rangle\,,
\ee
where ($m\,, n = 0\,, 1\,, \cdots\,, 2r$)
\be
    \bigl( \textbf{A}^{\phantom{\intercal}}_r \bigr)^{}_{mn} = (-1)^m \, \delta^{}_{m\,,\,2r-n}\,,
        \qquad%
        \text{\emph{i.e.},}
        \qquad%
    \textbf{A}^{\phantom{\intercal}}_r =
    \begin{pmatrix}
        \!\!0 & \cdots &\,\, 0 &\quad 0 &\quad 1\,\, \\[4pt]
        \!\!0 & \cdots &\,\, 0 &\quad -1 &\quad 0\,\, \\[4pt]
        \!\!0 & \cdots &\,\, 1 &\quad 0 &\quad 0\,\, \\[2pt]
        \!\!\vdots & 
        \iddots &\,\, \vdots &\quad \vdots &\quad \vdots\,\, \\[4pt]
        (-1)^{2r} & \cdots &\,\, 0 &\quad 0 &\quad 0\,\,
    \end{pmatrix}\,.
\ee
Using \eqref{eq:gsi0matrix} and the identity
\be
    \textbf{U}^\intercal_{2r+1} \, \textbf{A}^{\phantom{\intercal}}_r \, \textbf{U}^{\phantom{\intercal}}_{2r+1} = \textbf{A}^{\phantom{\intercal}}_{r} \,, 
\ee
we find $g \circ I^{\phantom{\intercal}}_r = I^{\phantom{\intercal}}_r$\,, \emph{i.e.}, $I^{}_r$ in Eq.~\eqref{eq:irsas} is SL($2\,,\mathbb{R}$) invariant.
Further note that Eq.~\eqref{eq:Im} forms a complete basis if the inner products between $s^{}_m$ do not accidentally vanish. 

To prove the above statement, we consider a general quadratic SL($2\,,\mathbb{R}$) invariant
\be \label{eq:ginvi}
    \mathscr{I} = \sum_{\ell\,,\, m} a^{}_{\ell\,,\,\ell+m} \, \Bigl\langle s^{}_\ell\,,  s^{}_{m} \Bigr\rangle  
\ee
that satisfies $g \circ \mathscr{I} = \mathscr{I}$.
Under the SL($2\,,\mathbb{R}$) group action, $g \circ \langle s_i\,, s_j \rangle$ is a degree-$(i + j)$ polynomial in $\kappa$\,. Suppose the highest degree among all the summands in Eq.~\eqref{eq:ginvi} is $h$\,. Collecting all summands in Eq.~\eqref{eq:ginvi} that transform into a degree-$h$ polynomial defines
\be \label{eq:ih}
    \mathscr{I}^{}_h = \sum_{m=0}^h a^{}_{m, \, h} \Bigl\langle s^{}_m\,,  s^{}_{h-m} \Bigr\rangle\,.
\ee
%
By construction, $g \circ \mathscr{I}_h$ is a degree-$h$ polynomial in $\kappa$\,. Using Eq.~\eqref{eq:gsi0}, we find
\be \label{eq:khpq}
    g \circ\mathscr{I}_h = \sum_{p=0}^h \sum_{q=0}^{h-p} (-\kappa)^{h-p-q} \, \Bigl\langle s^{}_p\,, \, s^{}_{q} \Bigr\rangle \sum_{m=0}^{h-p-q} \frac{a^{}_{m+p, \, h}}{m! \, (h-m-p-q)!}\,,
\ee
where a summand takes the form $\kappa^{h-p-q} \, \langle s^{}_p\,, \, s^{}_{q} \rangle$\,,
while a summand in $g \circ \bigl( \mathscr{I} - \mathscr{I}^{}_h \bigr)$ takes a different form $\kappa^u \, \langle s^{}_p\,, \, s^{}_{q} \rangle$\,, with $u + p + q < h$\,. Therefore, $\mathscr{I}_h$ has to be invariant on its own, \emph{i.e.}, $g \circ \mathscr{I}_h = \mathscr{I}_h$\,, which implies 
\be \label{eq:hpqa0}
    \sum_{m=0}^{h-p-q} \frac{a^{}_{m+p, \, h}}{m! \, (h-m-p-q)!} = 0\,, 
        \qquad%
    0 \leq p + q \leq h-1\,.
\ee
When $p+q=h-1$\,, we find $a_{p+1,\,h} = - a_{p,\,h}$ for $0 \leq p \leq h-1$\,. It then follows that 
\be \label{eq:amh}
    a^{}_{m,\,h} = (-1)^m \, a^{}_{0,\,h}\,,
        \qquad%
    0 \leq m \leq h\,.
\ee
The relations in Eq.~\eqref{eq:amh} solve Eq.~\eqref{eq:hpqa0}. 
Plugging Eq.~\eqref{eq:amh} into Eq.~\eqref{eq:ih}, we find
\be \label{eq:Ihr}
    \mathscr{I}_h = a^{}_{0,\,h} \sum_{m=0}^h (-1)^m \, \Bigl\langle s^{}_m\,,  s^{}_{h-m} \Bigr\rangle\,,
\ee
which is identically zero if $h$ is odd. When $h$ is even, Eq.~\eqref{eq:Ihr} is one of the basis invariants in Eq.~\eqref{eq:Im}. This procedure can be repeated for $\mathscr{I}-\mathscr{I}_h$ until we find zero, which implies
\be
    \mathscr{I} = \sum_{r=0}^{\lfloor h/2 \rfloor} a^{}_{0,2r} \, I^{}_{r}\,,
\ee
\emph{i.e.}, any SL($2\,,\mathbb{R}$) invariant $\mathscr{I}$ is a linear combination of the basis invariants in Eq.~\eqref{eq:Im}. 

In contrast, there is only one quadratic invariant associated with the GL($2\,,\mathbb{R}$) transformation \eqref{eq:gl2r} (instead of Eq.~\eqref{eq:gxy} that we have been considering) for a degree-$N$ binary form in classical invariant theory,
\be \label{eq:if2n}
    \CI^{}_N = \sum_{m=0}^{N-1} (-1)^m \, s^{}_m \, s^{}_{N-m-1}\,.
\ee
Note that the RHS of Eq.~\eqref{eq:if2n} vanishes identically when $N$ is even. 

For later application to IIB supergravity, we note the following subtlety: In the classification of invariants formed from the fields in an $N$-dimensional realization $\mathbf{S}^{}_N$ of SL($2\,,\mathbb{Z}$) in Eq.~\eqref{eq:defsu}, one might have to resort to higher-dimensional vectors. This is because an inner product \eqref{eq:Im} involving components in both $\mathbf{S}^{}_N$ and the higher-dimensional vectors may accidentally vanish. We will see an explicit example of this in the construction of NL IIB supergravity around Eq.~\eqref{eq:accvanish}.   

\section{Application to Non-Lorentzian IIB Supergravity} \label{sec:anl2bs}

Finally, we are ready to apply the mathematical machinery that we have developed in section~\ref{sec:prsl2rit} to NL IIB supergravity. We start with a dictionary of the notation for the reader. 
\begin{itemize}

\item

$\CO^{(i)}$ is a differential form of degree $i$\,.

\item

$\mathbf{S}^{(i)}_N$ is an $N$-dim. SL($2\,, \mathbb{R}$) realization whose components are degree-$i$ differential forms\,. 

\item

$s^{(i)}_m$ is a component in the vector $\mathbf{S}^{(i)}_N$\,, where $m$ labels its location in $\mathbf{S}^{(i)}_N$\,. 

\item

$I^{(i)}_r$ is a quadratic invariant containing inner products of the form $\bigl\langle s^{(i)}_m\,, s^{(i)}_{2r-m} \bigr\rangle$\,. 

\end{itemize}
In section~\ref{sec:sl2rinl}, we will also see the notation $I^{(i, \,p)}_r$ associated with $\bigl\langle s^{(i)}_m\,, s^{(i)}_{2r-m} \bigr\rangle_p$\,, where $p$ means that there are $p$ pairs of longitudinal frame indices being contracted in the inner product. 

\subsection{Supergravity Data as Binary Forms}

Using the invariant theory we developed in section~\ref{sec:gitbf}, we find that the information of NL IIB supergravity is encoded in the following two binary forms:
\begin{subequations} \label{eq:tpols}
\begin{align}
    \CP^{}_4 (x\,, \, y) & = 24 \, s^{(3)}_4 \, x^4 + 24 \, s^{(3)}_3 \, x^3 \, y + 12 \, s^{(3)}_2 \, x^2 \, y^2 + 4 \, s^{(3)}_1 \, x \, y^3 + s^{(3)}_0 \, y^4\,, \\[4pt]
    \CP^{}_2 (x\,, \, y) & = 2 \, s^{(5)}_2 \, x^2 + 2 \, s^{(5)}_1 \, x \, y + s^{(5)}_0 \, y^2\,,
\end{align}
\end{subequations}
where 
\be \label{eq:defxy}
    x = \CF^{(1)}\,,
        \qquad%
    y = - 3 \, \Bigl( \tfrac{1}{2} \, d\Phi + d\ln E \Bigr)\,,
\ee
with $E$ the determinant defined in \eqref{eq:defEdefF}. There are three polynomial realizations,
\be \label{eq:s253}
    \textbf{S}_2^{(1)} \!
    =
    \! \begin{pmatrix}
        x \\[4pt]
        y
    \end{pmatrix}\,,
        \qquad%
    \textbf{S}^{(3)}_5
    = 
    \begin{pmatrix}
        \CF^{(1)} \! \wedge \ell^{(2)} \\[4pt]
        \Gamma^{(3)} \\[4pt]
        \CF^{(3)} \\[4pt]
        \CH^{(3)} \\[4pt]
        \CA^{(3)}
    \end{pmatrix},
        \qquad%
    \textbf{S}^{(5)}_3 = 
    \begin{pmatrix}
        \CF^{(3)} \wedge \ell^{(2)} \\[4pt]
        \CH^{(3)} \wedge \ell^{(2)} \\[4pt]
        \CF^{(5)} 
    \end{pmatrix}.
\ee
Here, $\mathbf{S}^{(1)}_2$\,,  $\mathbf{S}^{(3)}_5$\,, and $\mathbf{S}^{(5)}_3$ contain one-, three-, and five-form fields, respectively. We have introduced a formal, commutative product between the differential forms in the polynomials in Eq.~\eqref{eq:tpols}, while preserving the associative and distributive properties.\,\footnote{The addition and product in Eq.~\eqref{eq:tpols} form a commutative ring.} We emphasize that this product is distinct from the wedge product between differential forms. For example, $x^4$ does not vanish even though $x$ is a one-form.

The mapping between the above ingredients and the field strengths in NL IIB supergravity introduced in section~\ref{eq:nlfs} is determined by
\begin{subequations} \label{eq:psi}
\begin{align}
    \CF^{(1)} & = 3 \, e^\Phi \, d C^{(0)}\,, 
        &%
    \Gamma^{(3)} & = d \ell^{(2)} - \tfrac{3}{2} \, d\Phi \wedge \ell^{(2)}\,, \label{eq:cg3} \\[4pt]
    \CF^{(5)} & = d C^{(4)} + C^{(2)} \wedge d B^{(2)}\,,
        &%
    \CF^{(3)} & = e^{\Phi/2} \, \bigl( dC^{(2)} + C^{(0)} \, dB^{(2)} \bigr)\,, \\[4pt]
        &%
        &
    \CH^{(3)} & = e^{-\Phi/2} \, d B^{(2)}\,.
\end{align}
\end{subequations}
Note $\CF^{(5)} = d\CC^{(4)} +\tfrac{1}{2} \bigl( dB^{(2)} \! \wedge C^{(2)} - dC^{(2)} \! \wedge B^{(2)} \bigr)$ in terms of $\CC^{(4)}$ in Eq.~\eqref{eq:defCC4}.
The three-form $\CA^{(3)}$ is related to the Lagrange multiplier $\CA^{(5)}$ in the NL IIB action \eqref{eq:NLIIB} and that imposes part of the self-duality constraints in Eq.~\eqref{eq:o3wl2}. The relation between $\CA^{(3)}$ and $\CA^{(5)}$ is given by
\begin{align} \label{eq:a5a3}
    \CA^{(5)} = \CA^{(3)}\wedge\ell^{(2)}\,.
\end{align}
From the point of view of polynomial realizations, $\CA^{(3)}$ is more fundamental than $\CA^{(5)}$\,: the new field $\CA^{(3)}$ is a part of the five-dimensional realization $\mathbf{S}^{(3)}_5$ in Eq.~\eqref{eq:s253}. Moreover, while $A^{(5)}$ is defined with a constraint $\CA^{(5)} \wedge \tau^A = 0$\,, $\CA^{(3)}$ is not constrained at all. This implies that $\CA^{(3)}$ contains the following St\"{u}ckelberg-type ambiguity: 
\be
    \CA^{(3)}\to\CA^{(3)} + \Theta_A^{(2)}\wedge\tau^A\,.
\ee
In the following, we will classify the invariants in NL IIB action in terms of $\CA^{(3)}$ instead of $\CA^{(5)}$\,. One may switch back to the notation in section~\ref{sec:nliibs} by using the condition \eqref{eq:a5a3}.

Finally, we note that the expressions of $y$ in Eq.~\eqref{eq:defxy} and $\Gamma^{(3)}$ in Eq.~\eqref{eq:cg3} are designed such that they both have a well-defined weight under the local dilatation transformation. We will further elaborate this in the next subsection.

\subsection{\texorpdfstring{Dilatation Weights and SL($2\,,\mathbb{R}$) Transformations}{Dilatation Weights and SL(2,R) Transformations}} \label{eq:dwslzt}

Now, we use the binary forms in Eq.~\eqref{eq:tpols} to generate the symmetry transformations of various field strengths. We first consider the local dilatation transformation parametrized by $\lambda^{}_\text{D} (x)$\,, which acts on an operator $\CO$ as in Eq.~\eqref{eq:dtrnsf}, \emph{i.e.}, $\CO \rightarrow \exp \bigl[\Delta (\CO) \, \lambda^{}_\text{D} \bigr] \, \CO$\,. 
Here, $\Delta(\CO)$ is the dilatation weight associated with the operator $\CO$ (see Eq.~\eqref{eq:dws}).
Note that the combinations $y \sim \frac{1}{2} \, d\Phi + d\ln E$ in Eq.~\eqref{eq:defxy} and $\Gamma^{(3)} = d\ell^{(2)} - \frac{3}{2} \, d\Phi \wedge \ell^{(2)}$ in Eq.~\eqref{eq:cg3} are required such that they have well-defined dilatation weights: the local dilatation transformation of $d\Phi$ is
\be
    d\Phi \rightarrow d\Phi + d\ln\lambda^{}_\text{D}   
\ee
where the shift $\ln \lambda^{}_\text{D}$ is canceled in the transformation of both $y$ and $\Gamma^{(3)}$\,. The dilatation weights of the components in the vectors $\mathbf{S}^{(1)}_2$\,, $\mathbf{S}^{(3)}_5$\,, and $\mathbf{S}^{(5)}_3$ are given by
\be \label{eq:dilw}
    \Delta \Bigl(s_m^{(1)}\Bigr) = 1 - m\,,
        \qquad%
    \Delta \Bigl(s_m^{(3)}\Bigr) = \frac{5}{2} - m\,,
        \qquad%
    \Delta \Bigl(s_m^{(5)}\Bigr) = 2 - m\,,
\ee
respectively. Any invariant term $E \, \bigl\langle s_i^{(n)}, s_j^{(n)} \bigr\rangle$ in the NL IIB Lagrangian has to have zero dilatation weight such that it is invariant under the local dilatation transformation. Here, $E$ is the measure defined in Eq.~\eqref{eq:defCC4} and its dilatation weight is $\Delta(E) = -1/2$\,. 

Next, we turn to the SL($2\,,\mathbb{R}$) transformations. The vectors $\mathbf{S}^{(1)}_2$, $\mathbf{S}^{(3)}_5$, and $\mathbf{S}^{(5)}_3$ form two-, five-, and three-dimensional polynomial realizations of SL($2\,,\mathbb{R}$), respectively. According to the transformation rule in Eq.~\eqref{eq:gsi0matrix} and the definitions in Eq.~\eqref{eq:s253}, we find the following SL($2\,,\mathbb{R}$) transformations of the IIB data:
\begin{subequations} \label{eq:unitrnsf}
\begin{align}
	\CF^{(1)} & \rightarrow \CF^{(1)}\,, \\[2pt]
	\Gamma^{(3)} & \rightarrow \Gamma^{(3)} - \kappa \, \CF^{(1)} \wedge \ell^{(2)}\,, \\[2pt]
	\CF^{(3)} & \rightarrow \CF^{(3)} - \kappa \, \Gamma^{(3)} + \tfrac{1}{2} \, \kappa^2 \, \CF^{(1)} \wedge \ell^{(2)}\,, \\[4pt]
	\CH^{(3)} & \rightarrow \CH^{(3)} - \kappa \, \CF^{(3)} + \tfrac{1}{2} \, \kappa^2 \, \Gamma^{(3)} - \tfrac{1}{3!} \, \kappa^3 \, \CF^{(1)} \wedge \ell^{(2)}\,, \\[4pt]
    \CA^{(3)} & \rightarrow \CA^{(3)} - \kappa \, \CH^{(3)} + \tfrac{1}{2} \, \kappa^2 \, \CF^{(3)} - \tfrac{1}{3!} \, \kappa^3 \, \Gamma^{(3)} + \tfrac{1}{4!} \, \kappa^4 \, \CF^{(1)} \wedge \ell^{(2)}\,,  \label{eq:a3trnsf} \\[4pt]
    \CF^{(5)} & \rightarrow \CF^{(5)} - \kappa \, \CH^{(3)} \wedge \ell^{(2)}+ \tfrac{1}{2} \, \kappa^2 \, \CF^{(3)} \wedge \ell^{(2)}\,.  
\end{align}
\end{subequations}
Supplemented with the definition of $\kappa$ in Eq.~\eqref{eq:kappadef} and the SL($2\,,\mathbb{R}$) transformations of $C^{(0)}$ and $\Phi$ in Eq.~\eqref{eq:e12slz}, which we transcribe below,
\be
    \kappa = \frac{\gamma \, e^{\Phi}}{\gamma \, C^{(0)} + \delta}\,,
        \qquad%
    e^{\Phi/2}
    \begin{pmatrix}
        C^{(0)} \\[2pt]
        1
    \end{pmatrix}
        \rightarrow
    e^{\Phi/2}
    \begin{pmatrix}
        \alpha &\quad \beta \\[2pt]
        \gamma &\quad \delta
    \end{pmatrix}
    \begin{pmatrix}
        C^{(0)} \\[2pt]
        1
    \end{pmatrix},
\ee
we are ready to classify all the quadratic SL($2\,, \mathbb{R}$) invariants in the NL IIB action. 

\subsection{\texorpdfstring{SL($2\,,\mathbb{R}$) Invariants in Non-Lorentzian IIB Supergravity}{SL(2,R) Invariants in Non-Lorentzian IIB Supergravity}} \label{sec:sl2rinl}

For now, we are interested in classifying the Lagrangian terms that are quadratic in spacetime derivatives and are invariant under both global SL($2\,,\mathbb{R}$) and local dilatation. As we have stressed in section~\ref{sec:symmetries}, the higher-form gauge, spacetime diffeomorphism, longitudinal Lorentz boost, and transverse rotation are easy to take care of. We will leave the string Galilei boost symmetry to section~\eqref{sec:EFT}. 

The SL($2\,,\mathbb{R}$) invariants have been classified in Eq.~\eqref{eq:Im}, which contains an abstract inner product $\bigl\langle s^{}_m\,, s^{}_{n} \bigr\rangle$\,. We define this inner product ``$\langle \cdot\,, \cdot \rangle$" between two differential forms of the same degree to be 
\be \label{eq:sisj}
    \Bigl\langle s^{(i)}_m\,, s^{(i)}_n \Bigr\rangle_{\!p}
        =
        \frac{1}{p! \, (i-p)!} \, \bigl(s^{}_m\bigr){}^{}_{A^{}_1\cdots A^{}_p \, A'_1 \cdots A'_{i-p}} \, \bigl(s_m\bigr)^{A_1\cdots A_p \, A'_1 \cdots A'_{i-p}}\,.
\ee
Here, $p$ counts how many indices in each $s^{(i)}_m$ are contracted with the inverse longitudinal Vielbein field $\tau^\mu{}_A$\,. Due to over-antisymmetrization, $0 \leq p \leq \text{min}\{i\,, 2\}$\,. Explicitly,
\be
    \bigl(s^{}_m\bigr){}^{}_{A_1\cdots A_p \, A'_1 \cdots A'_{i-p}} = \bigl( s^{}_m \bigr)^{}_{\mu_1 \cdots \mu_i} \, \tau^{\mu_1}{}^{}_{A_1} \, \cdots \, \tau^{\mu_p}{}^{}_{A_p} \, E^{\mu_{p+1}}{}^{}_{A'_1} \, \cdots \, E^{\mu_i}{}^{}_{A'_{i-p}}\,,
\ee
where the curved indices $\mu^{}_k$ in the differential form $s^{(i)}_m$ of degree $i$ are antisymmetrized. Therefore, a single invariant $I_r$ in Eq.~\eqref{eq:Im} gives
\be \label{eq:Iir}
	I^{(i)}_r = 
    \begin{pmatrix}
        I_r^{(i,0)} \\[6pt]
        I_r^{(i,1)} \\[6pt]
        I_r^{(i,2)}
    \end{pmatrix}
    =
    \sum_{m=0}^{2r} (-1)^m
    \begin{pmatrix}
        \frac{1}{i!} \, \bigl\langle s^{(i)}_m\,,  s^{(i)}_{2r-m} \bigr\rangle_0 \\[6pt]
        \frac{1}{(i-1)!} \, \bigl\langle s^{(i)}_m\,,  s^{(i)}_{2r-m} \bigr\rangle_1 \\[6pt]
        \frac{1}{2! \, (i-2)!} \, \bigl\langle s^{(i)}_m\,,  s^{(i)}_{2r-m} \bigr\rangle_2
    \end{pmatrix}.
\ee
In each $s^{(i)}_m$ contained in $I^{(i,\,p)}_r$, $i$ counts the total number of its indices and $p$ counts how many of the total $i$ indices are contracted with inverse longitudinal Vielbeine. The dilatation weights of the components of each invariant $I^{(i)}_r$ in Eq.~\eqref{eq:Iir} have been designated in Eq.~\eqref{eq:dilw}.  

The quadratic SL($2\,,\mathbb{R}$) invariant terms $I_r^{(i)}$ with well-defined dilatation weights have been classified in Eq.~\eqref{eq:if2n}. In terms of the inner product  defined in Eq.~\eqref{eq:sisj}, these invariants give rise to $I^{(i\,,p)}_r$ in Eq.~\eqref{eq:Iir}. The building blocks for the SL($2\,,\mathbb{R}$) invariants that are quadratic in spacetime derivatives in the NL IIB supergravity Lagrangian are therefore 
\be
    \CL^{}_\text{NL IIB} \sim \sum_{i\,, \, p\,, \, r} E \, I^{(i,\,p)}_r\,, 
\ee
where $E$ is the determinant in Eq.~\eqref{eq:defCC4} with the dilatation weight $\Delta(E) = -1/2$\,. The complete classification of all non-trivial quadratic invariants that are relevant to NL IIB supergravity and respect exact global SL($2\,,\mathbb{R}$) and local dilatation is given below:
\begin{subequations} \label{eq:sldinvs}
\begin{align}
    I_0^{(1,\,1)} & = \CF^{(1)}_A \, \CF^{(1)}_B \, \eta^{AB}\,, \label{eq:inv7} \\[4pt]
    I_1^{(3,\,2)} & = \Gamma^{(3)}_{ABA'} \, \Gamma^{(3)ABA'} - 2 \, \CF^{(1)}_{A'} \, \CF^{(3)}_{A'AB} \, \epsilon^{AB}\,, \label{eq:inv1} \\[6pt]
    I_2^{(3,\,1)} & = \tfrac{1}{2} \, \CF^{(3)}_{AA'B'} \, \CF^{(3)AA'B'} - \Gamma^{(3)}_{AA'B'} \, \CH^{(3)AA'B'}\,, \label{eq:inv2} \\[6pt]
    I_3^{(3,\,0)} & = \tfrac{1}{3!} \Bigl(- \CH^{(3)}_{A'B'C'} \, \CH^{(3)}_{A'B'C'} + 2 \, \CF^{(3)}_{A'B'C'} \, \CA^{(3)}_{A'B'C'} \Bigr)\,, \label{eq:inv3} \\[6pt]
    I_1^{(5,\,2)} & = \tfrac{1}{3!} \, \Bigl( \CH^{(3)}_{A'B'C'} \, \CH^{(3)}_{A'B'C'} + \CF^{(3)}_{A'B'C'} \, \CF^{(5)}_{A'B'C'AB} \, \epsilon^{AB} \Bigr)\,, \label{eq:inv5} \\[6pt]
    I_2^{(5,\,1)} & = \tfrac{1}{4!} \, \CF^{(5)}_{A'B'C'D'A} \, \CF^{(5)}_{A'B'C'D'}{}^A\,. \label{eq:inv6} 
\end{align}
\end{subequations}
See the first part of appendix~\ref{app:dsli} for a detailed derivation of Eq.~\eqref{eq:sldinvs}, where we will highlight the subtlety that was pointed out at the end of section~\ref{sec:cqi}.

Moreover, in NL IIB supergravity, any expressions formed by the Vielbeine $\tau^{}_\mu{}^A$ and $E^{}_\mu{}^{A'}$ together with their inverses are trivially invariant under SL($2\,,\mathbb{R}$), as the Vielbeine fields are SL($2\,,\mathbb{R}$) invariant in Einstein's frame. After imposing the local dilatation symmetry up to a total derivative, we find the following two Lagrangian terms that are trivally invariant under SL($2\,,\mathbb{R}$)\,:\,\footnote{The first term in Eq.~\eqref{eq:rtau} is invariant under the local dilatation symemtry up to a total derivative while the second term is an exact invariant.}
\begin{align} \label{eq:rtau}
    E \, \Bigl( R + \tfrac{8}{3} \,
    \tau^{}_{A'A}{}^A \, \tau^{}_{A'B}{}^{B} \Bigr)\,,
        \qquad%
    E \, \tau^{}_{A'\{AB\}} \, \tau^{A'\{AB\}} \,.
\end{align}
Here, the curvature scalar $R$ is defined in Eq.~\eqref{eq:expr} and $\tau_{A'A}{}^B = E^\mu{}_{\!A'} \, \tau^\nu{}_{\!A} \, \p_{[\mu} \tau_{\nu]}{}^B$. Furthermore, $\tau_{A'\{AB\}} \equiv \tau_{A'(AB)} - \tfrac{1}{2} \, \eta_{AB} \, \tau_{A'C}{}^C$ reads off the symmetric traceless part of the tensor Note that there is no cosmological constant term as it violates the local dilatation symmetry.

So far, we have classified the quadratic terms that are exactly invariant under SL($2\,,\mathbb{R}$). There is also a Chern-Simons term that is invariant only up to a total derivative. In the second half of appendix~\ref{app:dsli}, we construct this missing Chern-Simons term by formally treating the 10D NL IIB theory as the boundary of an 11D theory. This gives rise to the unique Chern-Simons term with zero dilatation weight, 
\be \label{eq:wzinv}
    I^{(10)}_\text{CS} = \CC^{(4)} \wedge \CH^{(3)} \wedge \CF^{(3)} - \CF^{(5)} \wedge \CA^{(3)} \wedge \ell^{(2)}\,,
\ee
which is invariant under SL($2\,,\mathbb{R}$) up to a total derivative.

\subsection{From Non-Lorentzian IIB Action to Non-Lorentzian Bootstrap}\label{sec:EFT}

In section \ref{sec:nrstrlim}, we have shown that the reparametrizations in Eq.~\eqref{eq:NLredefinitions} allow for a well-defined $\omega\to\infty$ limit of type IIB supergravity. This is the non-relativistic string limit that leads us to the action \eqref{eq:NLIIB}, which realizes the bosonic symmetries as detailed in section~\ref{sec:symmetries}. Now, we would like to turn the question around and ask: \emph{what is the effective field theory (EFT) that is invariant under all these bosonic symmetries?} 

In the Lorentzian case, the resulting bosonic EFT invariant under the higher-form gauge, spacetime Poincar\'{e}, and global SL($2\,,\mathbb{R}$) symmetry is not very illuminating. Focusing on the Lagrangian terms quadratic in spacetime derivatives, we find the EFT
\begin{align} \label{eq:hseft}
\begin{split}
    \hat{S}^{}_\text{EFT} = \frac{1}{16 \pi G^{}_\text{N}} & \int d^{10} x \, \hat{E} \, \biggl[ \hat{R} + \alpha^{}_1 \, \tr \Bigl( \partial_{\mu} \hat{\CM} \, \partial^{\mu} \hat{\CM}^{-1} \Bigr) + \alpha^{}_2 \, \hat{\CH}_{\mu\nu\rho}^{\intercal} \, \hat{\CM} \, \hat{\CH}^{\mu\nu\rho} \biggr] \\[4pt]
    + \frac{1}{16 \pi G^{}_\text{N}} & \int \biggl( \alpha^{}_3 \, \hat{F}^{(5)}\wedge \star \hat{F}^{(5)} + \alpha^{}_4 \, \hat{\CC}^{(4)} \wedge \hat{\CH}^{(3) \intercal} \wedge \epsilon \, \hat{\CH}^{(3)} \biggr)\,.
\end{split}
\end{align}
Moreover, besides the above quadratic invariants, there is also a cosmological constant term 
\be \label{eq:cc}
    \hat{S}^{}_\Lambda = -\frac{1}{8\pi G^{}_\text{N}} \int d^{10} x \, \hat{E} \, \Lambda\,,
\ee
which is also invariant under all the bosonic symmetries.
Not surprisingly, we still need the fermionic sector and have to impose local supersymmetry to fix all the free parameters $\alpha^{}_{i}$\,, $i = 1, \cdots, 4$ and the cosmological constant $\Lambda$ in the EFT \eqref{eq:hseft}, such that the Lorentzian IIB action \eqref{eq:relaction} is recovered. As is well known, after incorporating the fermions and imposing supersymmetry, the bosonic sector \eqref{eq:relaction} of the IIB supergravity action is recovered, where $\alpha^{}_i$'s in Eq.~\eqref{eq:hseft} are now fixed to be
\be \label{eq:alphavalue}
    \alpha^{}_1 = \frac{1}{4}\,,
        \qquad%
    \alpha^{}_2 = - \frac{1}{12}\,,
        \qquad%
    \alpha^{}_3 = \alpha^{}_4 = - \frac{1}{4}\,,
\ee
and the cosomological constant $\Lambda$ is set to zero.
Now, the only coupling constant in the theory is the gravitational constant $G_N$\,.
Moreover, requiring the invariance under supersymmetry also imposes the self-duality condition \eqref{eq:sdcf5}.

The same question regarding the bosonic EFT, but now in the NL corner, turns out to be more interesting. The non-relativistic string limit of the action \eqref{eq:hseft} is generically singular for arbitrary $\alpha^{}_i$'s, and one has to fine tune the $\alpha^{}_i$ couplings such that this stringy limit even makes sense. This fine tuning is a consequence of tuning the electric $B$-field to its critical value such that it cancels the string tension in the non-relativistic string limit, which is analogous to a BPS limit and must inherits some imprints of supersymmetry. This strongly suggests that imposing the bosonic symmetries in NL IIB theory must already lead to a much more constrained EFT compared to the Lorentzian case in Eq.~\eqref{eq:hseft}, without explicitly using supersymmetry. In other words, we expect fewer coupling constants in the NL EFT after imposing the bosonic symmetries detailed in section~\ref{sec:symmetries} but before imposing supersymmetry. It is therefore motivating to classify the NL EFT invariant under all the bosonic symmetries. If all the coupling constants other than the gravitational constant $G^{}_\text{N}$ can be fixed just using the bosonic symmetries, this would provide us with a powerful tool for classifying invariants in supergravity without even considering the more involved fermionic sector. We will see that this is indeed the case for NL IIB theory, at least at the lowest $\alpha'$ order. The impact of this study is not limited to the NL corner; we will show evidence that it is possible to ``bootstrap" the bosonic action in Lorentzian supergravity from the NL theory.

\paragraph{Non-Lorentzian IIB Action as a Bosonic EFT.} Just as in the Lorentzian case, the higher-form gauge symmetries are manifestly realized by requiring the gauge invariance of all the field strengths, and the spacetime diffeomorphisms are made manifest by appropriately contracting the curved indices $\mu = 0\,, 1\,, \cdots\,, 9$\,. Moreover, the local Lorentz boost symmetry in the 2D longitudinal sector and the local spatial rotation symmetry in the 8D transverse sector are made manifest by properly contracting the frame indices $A$ and $A'$\,. See section~\ref{sec:symmetries} for further details. In contrast, the remaining symmetries, namely, the local string Galilei boost, anisotropic dilatation symmetry, and the global SL($2\,,\mathbb{R}$) symmetry, are less manifest in the action \eqref{eq:NLIIB}. We have spent the bulk of this section on systematically formulating the polynomial realizations of the global \SLR\ and their interplay with the local anisotropic dilatation symmetry, culminating in the derivation of a complete list of invariants quadratic in spacetime derivatives in Eqs.~\eqref{eq:sldinvs}, \eqref{eq:rtau}, and \eqref{eq:wzinv}. 
At this point, the only remaining bosonic symmetry that has \emph{not yet} been imposed is the local string Galilei boost symmetry (see Eq.~\eqref{eq:sgbteb}). In the absence of this boost symmetry, the NL effective action quadratic in spacetime derivatives is
\begin{align} \label{eq:seft}
    S_{\text{EFT}} = \frac{1}{16 \pi G^{}_\text{N}}  & \int d^{10}x\,E\,\biggl( R + \tfrac{8}{3} \, \tau^{}_{A'A}{}^A \, \tau^{}_{A'B}{}^{B} + \beta^{}_1 \, \tau^{}_{A'\{AB\}} \, \tau^{A'\{AB\}} \biggr) \notag \\[4pt] 
    + \frac{1}{16 \pi G^{}_\text{N}} & \int d^{10} x \, E \, \biggl( \beta^{}_2\,  I_0^{(1,\,1)} + \beta^{}_3 \, I_1^{(3,\,2)} + \beta^{}_4 \, I_2^{(3,\,1)} + \beta^{}_5 \, I_3^{(3,\,0)} + \beta^{}_6 \, I_1^{(5,\,2)} + \beta^{}_7 \, I_2^{(5,\,1)} \biggr) \notag \\[4pt]
    + \frac{1}{16 \pi G^{}_\text{N}} & \int \beta^{}_{8} \, I^{(10)}_\text{CS}\,, 
\end{align}
where there are eight coupling constants $\beta^{}_i$\,, $i = 1\,, \cdots, 8$ in addition to the gravitational constant $G^{}_\text{N}$\,. We already noted in section~\eqref{sec:sl2rinl} that the NL analog of the cosmological constant term \eqref{eq:cc} is excluded due to the local dilatation symmetry \eqref{eq:dtrnsf}, which is a symmetry that is absent in the Lorentzian case. Fascinatingly, upon requiring that Eq.~\eqref{eq:seft} be invariant under the string Galilei symmetry \eqref{eq:sgbteb}, we find that all the $\beta^{}_i$'s are uniquely fixed. Namely,
\begin{subequations} \label{eq:bv}
\begin{align}
    \beta^{}_1 & = 0\,,
        &
    \beta^{}_2 & = - \frac{1}{18}\,,
        &
    \beta^{}_3 & = \frac{1}{12}\,,
        &
    \beta^{}_4 & = - \frac{1}{2}\,, \\[6pt]
    \beta^{}_5 & = \frac{1}{4}\,, 
        &
    \beta^{}_6 & = - \frac{1}{4}\,, 
        &
    \beta^{}_7 & = - \frac{1}{4}\,,
        &
    \beta^{}_8 & = - \frac{1}{2}\,.
\end{align}
\end{subequations}
The resulting NL EFT is
\be \label{eq:seftf}
\begin{split}
    S^{}_\text{EFT} &= \frac{1}{16 \pi G^{}_\text{N}} \,  \int d^{10}x\,E\,\biggl( R   + \tfrac{8}{3}\,\tau^{}_{A'A}{}^A \, \tau^{}_{A'B}{}^{B} \biggr) - \frac{1}{32 \pi G^{}_\text{N}} \, \int I^{(10)}_\text{CS} \\[4pt] 
    & \,\,\, - \frac{1}{64 \pi G^{}_\text{N}} \int d^{10} x \, E \, \biggl( \tfrac{2}{9}\,  I_0^{(1,\,1)} - \tfrac{1}{3} \, I_1^{(3,\,2)} + 2 \, I_2^{(3,\,1)} - I_3^{(3,\,0)} + I_1^{(5,\,2)} + I_2^{(5,\,1)} \biggr) \,,
\end{split}
\ee
where we have plugged the values of $\beta^{}_i$ in Eq.~\eqref{eq:bv} back into the action \eqref{eq:seft}. Recall that the expressions of $I^{(i,\,p)}_r$ and $I^{(10)}_\text{CS}$ are given in Eqs.~\eqref{eq:sldinvs} and \eqref{eq:wzinv}, respectively. Now, there is only a single coupling $G^{}_\text{N}$ in the theory. Using the relations in Eqs.~\eqref{eq:psi} and \eqref{eq:a5a3}, it is a straightforward exercise to show that the EFT \eqref{eq:seftf} is identical to the action principle \eqref{eq:NLIIB} that we have obtained from the non-relativistic string limit of the Lorentzian IIB action \eqref{eq:relaction}.

We therefore arrive at a remarkable conclusion: the bosonic sector of the NL IIB supergravity action can be constructed by only imposing the bosonic symmetries, without any explicit reference to the fermionic sector \emph{or} supersymmetry!\,\footnote{In other words, the action \eqref{eq:seftf} is an irreducible singlet under the bosonic symmetries.} 

Furthermore, it turns out that the NL EFT \eqref{eq:seftf} together with the non-relativsitic string limit already provide us with sufficient information to recover the bosonic part of Lorentzian IIB supergravity without explicitly resorting to the fermionic sector:
\begin{enumerate}[(1)]

\item

Requiring that the NL action \eqref{eq:seftf} arise from the non-relativistic string limit of the Lorentzian EFT action \eqref{eq:hseft} allows us to uniquely fix all the $\alpha^{}_i$ couplings precisely as in Eq.~\eqref{eq:alphavalue}. We explain a systematic way how this can be done. First, we match the $O(\omega^0)$ terms in the large $\omega$ expansion of the Lorentzian EFT \eqref{eq:hseft} with the associated terms in the NL EFT \eqref{eq:seftf}. It turns out that this procedure already fixes all the $\alpha^{}_i$ as in Eq.~\eqref{eq:alphavalue}. With these fixed $\alpha^{}_i$\,, as we have learned in section~\ref{sec:nrstrlim}, all the quadratic $\omega$ divergences in the Lorentzian EFT \eqref{eq:hseft} combine nicely into a complete square and give rise to the terms containing the Lagrange multiplier $\CA^{(3)}$ that match the associated ones in the NL EFT \eqref{eq:seft}.\,\footnote{It is interesting to note that, even before imposing the boost symmetry, matching the $O(\omega^0)$ terms already leads to the following constraints on both $\alpha^{}_i$ and $\beta^{}_i$\,: $\alpha^{}_1 = 1/4$\,, $\alpha^{}_2 = - 1 / 12$\,, $\alpha^{}_3 = \beta^{}_6$\,, $\alpha^{}_4 = \beta^{}_8 / 2$\,, $\beta^{}_6 = \beta^{}_7 = \beta^{}_5 - \frac{1}{2}$\,, and $\beta^{}_{i}$\,, $i = 1\,, \cdots, 4$ are fixed as in Eq.~\eqref{eq:bv}. Note that only $\beta^{}_6$ and $\beta^{}_8$ remain to be fixed by the string Galilei boost symmetry in the NL EFT.}

\item

As we have noted earlier, the analog of the cosmological constant term \eqref{eq:rtau} is forbidden in NL EFT \eqref{eq:seftf} due to the dilatation symmetry. By requiring the consistency of the non-relativistic string limit, the cosmological constant $\Lambda$ in Eq.~\eqref{eq:rtau} is also set to zero.

\item

The Lagrange multiplier $\CA^{(3)}$ in Eq.~\eqref{eq:seftf} (or, equivalently, $\CA^{(5)}$ in Eq.~\eqref{eq:NLIIB}) imposes the constraint \eqref{eq:transselfdual} on the three- and five-form field strengths. Lifting this constraint to the Lorentzian IIB theory, its covariantization gives rise to the self-duality condition \eqref{eq:sdcf5} in the Lorentzian theory. 

\end{enumerate}
In this way, we are able to derive the bosonic part of Lorentzian IIB supergravity action in Eq.~\eqref{eq:relaction} together with the self-duality condition, without considering the fermions at all!

\paragraph{A Non-Lorentzian Bootstrap.} The above procedure provides a potentially powerful method for determining the bosonic part of Lorentzian IIB supergravity action purely using bosonic symmetries, order by order in the Regge slope coupling $\alpha'$\,. A subtlety to note is that, when higher-curvature corrections are included, only the global SL($2\,,\mathbb{Z}$) (instead of SL($2\,,\mathbb{R}$)) symmetry can be restored \cite{Green:1997tv}.\,\footnote{We would like to thank Axel Kleinschmidt for pointing this out.} However, this subtlety does not affect any of our arguments here. We summarize the detailed steps below:
\begin{enumerate}[(1)]

\item \label{item:step1}

We start with the classification of a bosonic, NL EFT by requiring its invariance under all the bosonic symmetries in NL IIB supergravity, which in particular include the local dilatation, string Galilei boost, and global SL($2\,,\mathbb{R}$) (or SL($2\,,\mathbb{Z}$)) symmetry. 

\item \label{item:step2}

In parallel, we derive a bosonic, Lorentzian EFT by requiring the theory to be invariant under all the bosonic symmetries in Lorentzian IIB supergravity, which in particular include the SL($2\,,\mathbb{R}$) (or SL($2\,,\mathbb{Z}$)) symmetry. 

\item

We require that the NL EFT found in step (\ref{item:step1}) arise from the non-relativistic string limit of Lorentzian EFT found in step (\ref{item:step2}). This step is supposed to further constrain the coupling constants in the latter theory and determines the final Lorentzian EFT. 

\end{enumerate}
We emphasize for another time that the above method does not explicitly involve any fermion or supersymmetry.
In this subsection, we have seen that the above method works extremely well for determining the lowest-order bosonic terms in Lorentzian IIB supergravity. It would be fascinating if the same method can also be applied to extract higher-order bosonic terms in Lorentzian IIB supergravity from the classification of the NL IIB action. These lines of exploration may eventually lead to the idea of \emph{non-Lorentzian bootstrap}, where bosonic quantities in Lorentzian IIB supergravity are constrained using bosonic symmetries in a smaller, NL theory, by requiring that the latter be embeddable within the former theory.  

\vspace{3mm}

Before ending this section, we present a few more comments. First, we emphasize the power of the string Galilei boost symmetry to fix almost all the coupling constants in Eq.~\eqref{eq:seft}. In fact, without imposing the global SL($2\,,\mathbb{R}$) but requiring all the other bosonic symmetries including the string Galilei boost, the NL EFT can already be fixed up to two coupling constants $G^{}_\text{N}$ and $\beta$\,, where $G^{}_\text{N}$ is the gravitational constant and $\beta$ the coupling between the NS-NS and RR sectors. The resulting NL IIB EFT without the SL($2\,,\mathbb{R}$) symmetry takes the following form:
\be \label{eq:sbeta}
    S^{}_\beta = \frac{1}{G_\text{N}} \, \Bigl( S^{}_\text{NS} + \beta \, S^{}_\text{R} \Bigr)\,. 
\ee
Here, $S_\text{NS}$ contains all the NS-NS terms, while $S_\text{R}$ contains all the RR terms and the term involving the Lagrange multiplier $\CA^{(3)}$\,. It has been shown in \cite{Bergshoeff:2021bmc} that $S_\text{NS}$ is invariant under the local boost and dilatation symmetry, and we also find that $S_\text{R}$ is invariant under these symmetries as well. Imposing the global SL($2\,, \mathbb{R}$) symmetry further fixes the coupling $\beta$ to be a constant value, such that the effective action \eqref{eq:sbeta} becomes identical to the NL IIB supergravity action \eqref{eq:seftf}. 

In contrast, in NL IIA theory, a similar splitting between the NS-NS and RR terms as in Eq.~\eqref{eq:sbeta} also takes place, but there is \emph{no} extra global SL($2\,,\mathbb{R}$) that allows us to further fix the coupling $\beta$ (unless T-duality between 9D IIA and IIB is taken into account). Moreover, there is also a cosmological constant that is not necessarily zero in the IIA case, which may lead to a NL version of massive IIA supergravity \cite{Romans:1985tz}.

Note that the key point of our proposed non-Lorentzian bootstrap is not so much that the symmetries that we use to fix the bosonic part of the NL IIB supergravity action are bosonic symmetries. The point is that there are other symmetries, such as supersymmetry, that turn out not to constrain the form of the bosonic part of the action any further. In fact there are more symmetries, such as the $O(10,10)$ symmetry group made manifest through Double Field Theory (see \cite{Jeon:2012hp, Ko:2015rha} for context in IIA/IIB supergravity and nonrelativistic string theory), and the U-duality group of Exceptional Field Theory \cite{Berman:2019izh}. Certainly, in the latter, the bosonic symmetries fix the NL IIB action, but this is also a much larger group of symmetries than those we have used in this paper.

\section{Outlook} \label{sec:outlook}

Beyond determining the \SLR\ symmetry structure of NL IIB supergravity, there are several natural directions to pursue. We leave the following points for future investigations.

\paragraph{Equations of Motion.}

Apart from the realization of the SL($2\,, \mathbb{R}$) symmetry, other symmetries that are realized non-trivially in NL IIB supergravity are the boost symmetry and an emergent anisotropic dilatation symmetry. The dilatation symmetry was found before in the context of both string sigma models \cite{Bergshoeff:2018yvt} and 10D $\CN=1$ supergravity \cite{Bergshoeff:2021tfn}. They have the effect that the Poisson equation describing the dynamics of the string Newton-Cartan supergravity theory does \emph{not} follow from a variation of the NL action.\,\footnote{It is not surprising that the Poisson equation is not captured by the action principle in Newton-Cartan-like theories. In fact, in Newtonian gravity, it has long been an open question to construct an action principle that captures the geometrized Poisson equation. There has been recent progress on deriving the Poisson equation in Newton-Cartan gravity by extending the field content to the so-called type II Newton-Cartan geometry \cite{Hansen:2019pkl}. Also see \cite{Gallegos:2020egk} for later generalizations to string Newton-Cartan gravity.} This Poisson equation arises from the non-relativistic limit of the equations of motion \cite{Bergshoeff:2021bmc}, or, intrinsically, from the beta-functions of the worldsheet sigma model describing non-relativistic string theory \cite{Gomis:2019zyu, Gallegos:2019icg}. The NL action should therefore be viewed as a pseudo-action that gives rise to most but not all equations of motion. Additional contributions to the Poisson equation from the Ramond-Ramond sector are also expected. It would be interesting to see how the \SLR\, symmetry is realized on the equations of motion. In particular, it is natural to ask how the field equations of motion, including the Poisson equation, fit into the framework of the polynomial realization of SL($2\,,\mathbb{R}$)\,.

There is also an interesting aspect regarding the equations of motion in Lorentzian IIB supergravity. It is well known that there is a self-duality condition of the RR five-form field-strength in Lorentzian IIB supergravity. This condition, even though it constrains the bosonic contents, does not follow from the bosonic action principle but only arises as an equation of motion in the full supergravity theory. In section~\ref{sec:nrstrlim}, we showed that, after taking the non-relativistic string limit, this self-duality condition gives rise to two separate conditions. These resulting conditions together form a reducible but indecomposable representation under the boost transformations. Remarkably, in contrast to the case in Lorentzian IIB theory, one of these conditions (but not both) does follow from the NL IIB action. This condition is precisely the equation of motion imposed by the Lagrange multiplier $\CA^{(5)}$ that we discussed in the bulk of the paper. This new field is an indispensable part of the IIB supergravity action and of the polynomial realization of SL($2\,,\mathbb{R}$). It is therefore important to further understand the role of $\CA^{(5)}$ as part of the supermultiplet in NL IIB supergravity.

\paragraph{Supersymmetry and Torsional Constraints.}

It is natural to extend the results of this paper to a supersymmetric theory, which entails adding fermionic fields along the lines shown for minimal supergravity in \cite{Bergshoeff:2021tfn}. There, it was also shown that the supersymmetric multiplet exists only if we impose non-trivial constraints on the intrinsic torsion associated with the longitudinal Vielbein field $\tau_\mu{}^A$\,. Since the IIB multiplet has maximal supersymmetry, further constraints are expected.\footnote{We were informed by Luca Romano that work on a NL IIB SUGRA including fermions is in progress \cite{Romano:2023xxx}.} This expectation is also supported by a study of the eleven-dimensional supermultiplet \cite{11dsugra}. 

Since the torsional constraints play a fundamental role in NL supergravity, it is desirable to understand them from the worldsheet perspective. Two different interacting, renormalizable string sigma models describing bosonic non-relativistic strings have been constructed from symmetry principles \cite{Yan:2021lbe}. Both the symmetry algebras involved in the construction of the string sigma models require certain non-central extensions of the string Galilei algebras:
\begin{itemize}

\item

The first choice requires that the string Galilei boosts and the transverse translations commute into a new generator $Z^{}_{\!A}$\,, whose realization on the worldsheet imposes the torsional constraint $D^{}_{[\mu} \tau^{}_{\nu]}{}^A = 0$\,, where $D^{}_{\mu}$ is covariantized with respect to the frame index $A$ \cite{Andringa:2012uz, Bergshoeff:2018yvt}. This constraint is expected to arise from requiring the self-consistency of $\CN=2$ supersymmetry transformations in IIA or IIB supergravity \cite{11dsugra}. 

\item

The second choice breaks half of the $Z^{}_{\!A}$ in a lightlike coordinate, whose realization on the worldsheet leads to the torsional constraint $\tau^{}_{[\mu}{}^- \p^{}_\nu \tau^{}_{\rho]}{}^- = 0$ \cite{Yan:2021lbe}. Here, the superscript ``$-$" is the lightlike index. This constraint also arises from requiring the self-consistency of $\CN=1$ supersymmetry transformations in 10D minimal supergravity \cite{Bergshoeff:2021tfn}. 

\end{itemize}
Both the torsional constraints are crucial for maintaining the renormalizability of the worldsheet quantum field theory such that the theory does \emph{not} generate extra counterterms at higher loop orders, which would have driven the theory towards the sigma models describing relativistic string theory \cite{Gomis:2019zyu, Yan:2019xsf, Yan:2021lbe}. 

It might be puzzling why there exist two self-consistent, bosonic non-relativistic string theories defined by different symmetry principles that lead to different torsional constraints. This is rather distinct from the relativistic case, where there is a unique bosonic string theory and there is \emph{a priori} no constraint on the target-space geometry. The observations we collected above strongly suggest a natural explanation in the context of non-relativistic superstrings: the bosonic non-relativistic string theory with the $Z^{}_A$ symmetry should be the bosonic part of non-relativistic superstring theories with $\CN=2$ supersymmetry, while the other one with the halved $Z^{}_A$ symmetry should correspond to non-relativistic superstrings with $\CN=1$ supersymmetry. The associated torsional constraints are also expected to play an important role in DLCQ string/M-theory and Matrix theory, in view of their duality relations to non-relativistic string/M-theory \cite{Gomis:2000bd,Danielsson:2000gi,Bergshoeff:2018yvt}. 

\paragraph{Higher-Curvature Corrections.}

Remarkably, the bosonic sector of NL IIB supergravity can be viewed as a bosonic effective field theory on its own, with the action being uniquely determined by the bosonic symmetries, without explicitly considering the fermionic sector. This is different from the Lorentzian case. 
It would be interesting to see whether the bosonic symmetries can also constrain the higher-curvature terms and thereby provide a systematic way of classifying the $\alpha'$-corrections in NL supergravity. Furthermore, requiring that the NL theory arise from the non-relativistic string limit of Lorentzian supergravity as in this paper, this could, in turn, serve as a tool to constrain the $\alpha'$-corrections in Lorentzian supergravity without considering fermions. See an example at the lowest $\alpha'$ order in IIB supergravity in section~\eqref{sec:EFT}, where the full bosonic sector including the self-duality constraint in Lorentzian IIB supergravity is recovered from the NL corner. It is valuable to further develop this novel idea of non-Lorentzian bootstrap in the future, probably also with generalizations beyond IIB supergravity. 

\paragraph{Non-Lorentzian Holography.}

One of the main motivations for investigating  NL IIB supergravity and its symmetries is because it allows us to study its half-supersymmetric D-brane solutions. Starting from the half-supersymmetric fundamental string solution  that we constructed in \cite{Bergshoeff:2022pzk} and using the duality symmetries, one is able to construct all the half-supersymmetric Dp-brane solutions for odd $p=1,3,5,7,9$ in NL IIB supergravity. The NL D7- and D9-branes should couple to the  NL version of the RR eight- and ten-form potentials in generalized IIB supergravity \cite{Bergshoeff:2005ac}. The zero-mode dynamics of these NL Dp-brane solutions should correspond to the worldvolume actions in non-relativistic string theory, which are derived from the worldsheet perspective in \cite{Gomis:2020fui} and then extended in \cite{Ebert:2021mfu} to include the RR potentials. These solutions can be used as a convenient starting point to speculate about a top-down realization of holography with NL supergravity in the bulk. We hope to come back to this possibility in a future work.

\paragraph{Other Ramond-Ramond Potentials.}

In this work, we considered a minimal formulation of IIB supergravity. There are further higher-form RR potentials that couple to a doublet of five-branes, a triplet of seven-branes, and a quadruplet of nine-branes \cite{Bergshoeff:2005ac}. 
It would be interesting to see how the \SLR\ transformations of the NL versions of these higher-form RR potentials follow from the general formalism developed in this work.

Finally, in parallel to this work on NL IIB supergravity, it would be interesting to also consider the non-relativistic limit of Lorentzian IIA supergravity, where the reparametrization of the RR fields is in form the same as Eq.~\eqref{eq:rpcq} but with $q=1,3$ (instead of $q=0,2,4$). In the same spirit of our discussion from the previous paragraph, one may also consider other higher odd-form RR potentials. Na\"{i}vely, the non-relativistic string limit of the Lorentzian IIA action leads to a divergence in the NL IIA action, in a similar way as in Eq.~\eqref{eq:S2orig}. Just like in the IIB case, this divergence can be treated by performing a Hubbard-Stratanovich transformation and eventually leads to a well-defined NL IIA action with an extra Lagrange multiplier imposing certain constraints. Related discussions on NL IIA supergravity have appeared in \cite{Blair:2021waq}, albeit using a different approach. 

\acknowledgments

We would like to thank Chris Blair, Niels Obers, Gerben Oling, and Luca Romano for useful discussions. In particular, we thank Stephen Ebert for valuable discussions and comments on a draft of the paper. We would also like to thank the organizers of the workshop on \emph{Beyond Lorentzian Geometry II} at ICMS, Edinburgh in February 2023 and the workshop on \emph{Non-Relativistic Strings and Beyond} at Nordita, Stockholm in May 2023, where this work was presented and completed. 
K.T.G. has received funding from the European Union’s Horizon 2020 research and innovation programme under the Marie Sk\l odowska-Curie grant agreement No 101024967.  Z.Y. is
supported by the European Union’s Horizon 2020 research and innovation programme under
the Marie Sk\l odowska-Curie grant agreement No 31003710. U.Z. is supported by TUBITAK - 2218 National Postdoctoral Research
Fellowship Program with grant number 118C512. Nordita is supported in part by NordForsk.

\newpage

\appendix

\section{Non-Lorentzian Scalar Curvature} \label{sec:Ricci}

In section~\ref{sec:nrstrlim}, we have shown that the redefinitions \eqref{eq:NLredefinitions} lead to an expansion of the IIB action of the form \eqref{eq:somegaexp}. This calculation is straightforward once one finds the correct expansion of the Ricci scalar which takes the form 
\begin{align}
    \hat{R} = -\omega^{5/2}\,\tau^{}_{A'B'A} \, \tau^{A'B'A} + \omega^{1/2}\,R + \mathcal{O}(\omega^{-3/2})\,.
\end{align}
The leading order term cancels against contributions coming from the Kalb-Ramond field, leaving $R$ as the leading order. In order to give an explicit expression for the expansion of the Ricci scalar it is actually more useful to expand in the metric formalism where $\tau_{\mu\nu} = \eta_{AB}\tau_\mu{}^A\tau_\nu{}^B$, $E_{\mu\nu}=\delta_{A'B'} E_\mu{}^{A'} E_\nu{}^{B'}$, $\tau^{\mu\nu} = \eta^{AB}\tau^\mu{}_A\tau^\nu{}_B$, and $E^{\mu\nu} = \delta^{A'B'}E^\mu{}_{A'} E^\nu{}_{B'}$. Then 
\begin{align} \label{eq:expr}
    R &= \tau^{\mu\nu}\big(\partial_\rho X^\rho_{\mu\nu} - \partial_\mu X^\rho_{\rho\nu} + X^\rho_{\rho\sigma}Y^\sigma_{\mu\nu} - X^\rho_{\mu\sigma}Y^\sigma_{\rho\nu}\big) \notag\\[4pt]
    &\quad+ E^{\mu\nu}\,\big(\partial_\rho Y^\rho_{\mu\nu} - \partial_\mu Y^\rho_{\rho\nu} + Y^\rho_{\rho\sigma}Y^\sigma_{\mu\nu} - Y^\rho_{\mu\sigma}Y^\sigma_{\rho\nu} +X^\rho_{\rho\sigma}Z^\sigma_{\mu\nu} - X^\rho_{\mu\sigma}Z^\sigma_{\rho\nu}\big)\,,
\end{align}
where
\begin{subequations}
    \begin{align}
    X^\rho_{\mu\nu} &= E^{\rho\sigma}\,\big(\partial_{(\mu}\tau_{\nu)\sigma} - \tfrac12\,\partial_\sigma\tau_{\mu\nu}\big)\,,\\
    %
    Y^\rho_{\mu\nu} &= \tau^{\rho\sigma}\,\big(\partial_{(\mu}\tau_{\nu)\sigma} - \tfrac12\,\partial_\sigma\tau_{\mu\nu}\big)+ E^{\rho\sigma}\,\big(\partial_{(\mu}E_{\nu)\sigma} - \tfrac12\,\partial_\sigma E_{\mu\nu}\big)\,,\\[4pt]
    Z^\rho_{\mu\nu} &= \tau^{\rho\sigma}\,\big(\partial_{(\mu}E_{\nu)\sigma} - \tfrac12\,\partial_\sigma E_{\mu\nu}\big)\,.
    \end{align}
\end{subequations}
The expression $R$ is manifestly invariant under \SLR\ transformations and has global dilatation weight $\Delta(R) = 1/2$\,. It is, however, not transforming covariantly under Galilean boosts and local dilatations
\begin{align}
    \delta R = -2\,\tau^{A'B'}{}^{}_{\!A} \Bigl(D^{}_{A'} \lambda^A{}^{}_{B'} + 2\,\lambda^{}_{BA'} \, \tau^{}_{B'}{}^{(AB)} \Bigr) - 4\,\tau_{A'A}{}^A \, \partial^{A'}\lambda_D
\end{align}
which follows from the fact that $R$ appears at subleading order in the expansion of the relativistic Ricci scalar. The derivative $D_\mu \lambda^{AA'}$ contains the $\mathrm{SO}(1,1)\times \mathrm{SO}(8)$ spin connections 
\be
    \omega_\mu{}^{AB} = \tau^{\nu B} \Bigl(\partial_\mu \tau_\nu{}^A - Y_{\mu\nu}^\rho\tau_\rho^A \Bigr)\,, 
        \qquad%
    \omega_\mu{}^{A'B'} = E^{\nu B'} \Bigl(\partial_\mu E_\nu{}^{A'} - Y_{\mu\nu}^\rho E_\rho{}^{A'}\Bigr)\,.
\ee
For manipulations involving the scalar curvature it is often useful to use the following identity 
\be
    \partial_\mu\big(E\,E^\mu{}_{A'}\big) = E\,\Bigl( E^\mu{}_{B'} \, \omega_{\mu A'}{}^{B'} + 2\,\tau_{A'A}{}^A\Bigr).
\ee

In \cite{Bergshoeff:2021bmc}, an improved curvature scalar $R(J)$ has been defined, which transforms covariantly under boosts. This is achieved by adding appropriate Kalb-Ramond terms in the definition. The relation between the two scalars can straightforwardly be worked out by using the relation between the string and Einstein expansions
\begin{equation}
    R^{}_{{\rm Einstein}} = e^{\Phi / 2} \biggl( R^{}_{{\rm string}} + \frac{9}{2} \, D^{A'} \partial_{A'} \Phi - \frac{9}{2} \, \partial_{A'} \Phi \, \partial^{A'} \Phi + 9\,\tau_{A'A}{}^A\partial^{A'}\Phi\biggr)\,,
\end{equation}
where $D^{}_{\!A'}$ contains an $\mathrm{SO}(8)$ connection that can be obtained by replacing the Einstein frame fields in the previously defined connections by string frame fields. 

\section{Expansion of the IIB Supergravity Action} \label{app:eiibsa}

In this appendix, we present details of the expansion of the Lorentzian IIB supergravity action \eqref{eq:relaction} with respect to the parameter $\omega$ introduced in Eq.~\eqref{eq:NLredefinitions}. We start by separating the Lorentzian action \eqref{eq:relaction} into different manifestly SL$(2\,,\mathbb{R})$-invariant terms as
\begin{equation} \label{eq:relaction2}
    \hat{S} = S^{}_{{\rm EH}} + S_{\hat{\CM}} + S_{\hat{\CH}} + S_{\hat{F}^{(5)}} + S^{}_{{\rm CS}}\,,
\end{equation}
where
\begin{subequations} \label{eq:relactionterms}
\begin{align}
    S^{}_{{\rm EH}} &= \frac{1}{16\pi G^{}_\text{N}} \int d^{10} x \, \hat{E} \hat{R}\,, \\[4pt]
    S_{\hat{\CM}} &= \frac{1}{16\pi G^{}_\text{N}} \int d^{10} x \, \hat{E} \, \tr \Bigl( \tfrac{1}{4} \, \partial_{\mu} \hat{\CM} \, \partial^{\mu} \hat{\CM}^{-1} \Bigr)\,, \\[4pt]
    S_{\hat{\CH}} &= \frac{1}{16\pi G^{}_\text{N}} \int d^{10} x \, \hat{E} \, \Bigl( - \tfrac{1}{12}  \, \hat{\CH}_{\mu\nu\rho}^{\intercal} \, \hat{\CM} \, \hat{\CH}^{\mu\nu\rho} \Bigr)\,, \\[4pt]
    S_{\hat{F}^{(5)}} &= \frac{1}{16\pi G^{}_\text{N}} \int \tfrac{1}{4} \, \hat{F}^{(5)}\wedge \star \hat{F}^{(5)}\,, \\[4pt]
    S_{{\rm CS}} &= \frac{1}{16\pi G^{}_\text{N}} \int \Bigl( - \tfrac{1}{4} \, \hat{\CC}^{(4)} \wedge \hat{\CH}^{(3)}{}^\intercal_{\phantom{I}} \wedge \epsilon \, \hat{\CH}^{(3)} \Bigr)\,,
\end{align}
\end{subequations}
Later in \eqref{eq:somegaexp}, we rewrote the Lorentzian action \eqref{eq:relaction2} using Eq.~\eqref{eq:NLredefinitions} and expanded it with respect to a large $\omega$\,, \emph{i.e.}, 
\begin{align} \label{eq:somegaexp2}
    \hat{S} = \omega^2 \qdiv{S} + \stackrel{(0)}{S} + O(\omega^{-2})\,.
\end{align}
The $\omega^2$ terms are given in Eq.~\eqref{eq:S2orig}, which combine the following contributions from all the different terms in Eq.~\eqref{eq:relactionterms}:
\begin{subequations} \label{eq:qdiv}
\begin{align}
    \stackrel{(2)}{S}^{}_{\!{\rm EH}} &= \frac{1}{16\pi G^{}_\text{N}} \int d^{10} x \, E \, \biggl( - \tau^{}_{A'B'A} \tau^{A'B'A} \biggr)\,, \\[4pt]
    \stackrel{(2)}{S}_{\!\hat{\CH}} &= \frac{1}{16\pi G^{}_\text{N}} \int d^{10} x \, E \, \biggl( \tau^{}_{A'B'A} \tau^{A'B'A} + \tfrac{1}{2} \, e^{2 \Phi} \, F^{}_{A'} \, F^{A'} - \tfrac{1}{2 \cdot 3!} \, e^{\Phi} \, F^{}_{A'B'C'} \, F^{A'B'C'} \biggr) \,, \\[4pt]
    \stackrel{(2)}{S}_{\!\hat{\CM}} &= \frac{1}{16\pi G^{}_\text{N}} \int d^{10} x \, E \, \biggl( - \tfrac{1}{2} \, e^{2 \Phi} F_{A'} F^{A'} \biggr) \,, \\[4pt]
    \stackrel{(2)}{S}_{\!\hat{F}_5} &= \frac{1}{16\pi G^{}_\text{N}} \int d^{10} x \, E \, \biggl( \tfrac{1}{4!} \, e^{\Phi} \, F^{}_{A'B'C'} \, F^{A'B'C'} - \tfrac{1}{4 \cdot 5!} \, F_{A_1' \cdots A_5'} \, F^{A_1' \cdots A_5'} \biggr) \,, \\[4pt]
    \stackrel{(2)}{S}_{\!{\rm CS}} &= \frac{1}{16\pi G^{}_\text{N}} \int \biggl( - \tfrac{1}{2} \, e^{\Phi / 2}  \, F^{(5)} \wedge F^{(3)} \wedge \ell^{(2)} \biggr)\,,
\end{align}
\end{subequations}
where various integrations by parts (ignoring boundary terms) had to be performed in order to bring the Chern-Simons term in the form given above. Summing over the expressions in Eq.~\eqref{eq:qdiv} gives Eq.~\eqref{eq:S2orig}.
Furthermore, the $O(\omega^0)$ terms in Eq.~\eqref{eq:somegaexp2} receive the following contributions from Eq.~\eqref{eq:relaction2}: 
\begin{subequations} \label{eq:0so}
\begin{align}
    \stackrel{(0)}{S}_{{\rm EH}} &= \frac{1}{16\pi G^{}_\text{N}} \int d^{10} x \, E \, R\,, \\[4pt]
    \stackrel{(0)}{S}_{\hat{\CH}} &= \frac{1}{16\pi G^{}_\text{N}} \int d^{10} x \, E \, \biggl( 2 \, \tau^{}_{A'A}{}^{A} \, \tau^{A'B}{}_{B} + \tau^{A'A}{}_{A} \, \partial^{}_{A'} \Phi + \tfrac{1}{8} \, \partial_{A'} \Phi \, \partial^{A'} \Phi \notag \\[2pt]
    & \hspace{4cm} + e^{- \frac{1}{2}\Phi} \, \epsilon^{}_{AB} \, \tau^{}_{A'B'}{}^{A} \, H^{A'B'B} - \tfrac{1}{2 \cdot 3!} \, e^{- \Phi} H_{A'B'C'} H^{A'B'C'}  \notag \\[4pt]
    & \hspace{4cm} - \tfrac{1}{2} \, e^{\frac{3}{2} \Phi} \, F^{A'} F^{}_{A'AB} \, \epsilon^{AB} - \tfrac{1}{4} \, e^{\Phi} F_{A'B'A} F^{A'B'A}  \biggr)\,, \\[4pt]
    \stackrel{(0)}{S}_{\hat{\CM}} &= \frac{1}{16\pi G^{}_\text{N}} \int d^{10} x \, E \, \biggl( - \tfrac{1}{2} \, e^{2 \Phi} \, F^{}_A \, F^A - \tfrac{1}{2} \, \partial^{}_{A'} \Phi \, \partial^{A'} \Phi \biggr)\,, \\[4pt]
    \stackrel{(0)}{S}_{\hat{F}_5} &= \frac{1}{16\pi G^{}_\text{N}} \int d^{10} x \, \frac{E}{4!} \, \biggl( - e^{\frac{1}{2}\Phi} \, F^{A'B'C'} \, F^{}_{A'B'C'AB} \epsilon^{AB} - \tfrac{1}{4} \, F^{}_{A_1' \cdots A_4' A} \, F^{A_1' \cdots A_4'A} \biggr)\,, \\[4pt]
    \stackrel{(0)}{S}_{{\rm CS}} &= \frac{1}{16\pi G^{}_\text{N}} \, \int \biggl( - \tfrac{1}{2} \, \CC^{(4)} \wedge H^{(3)} \wedge F^{(3)} \biggr)\,,
\end{align}
\end{subequations}
summing over which gives
\begin{align}\label{eq:NLIIB2}
   \stackrel{(0)}{S} &= \stackrel{(0)}{S}^{}_{\!{\rm EH}} + \stackrel{(0)}{S}_{\!\hat{\CH}} + \stackrel{(0)}{S}_{\!\hat{\CM}} + \stackrel{(0)}{S}_{\!\hat{F}_5} + \stackrel{(0)}{S}^{}_{\!{\rm CS}}\,.
\end{align}
This is what we have recorded in Eq.~\eqref{eq:(0)S}. 

\section{\texorpdfstring{SL($2\,,\mathbb{R}$) Invariants in Non-Lorentzian IIB Supergravity}{Derivation of SL(2,R) Invariants in Non-Lorentzian IIB Supergravity}} \label{app:dsli}

This appendix contains some supplementary materials for section~\ref{sec:sl2rinl}. 

We first provide more details for how Eq.~\eqref{eq:sldinvs} is derived. Using Eqs.~\eqref{eq:dilw} and \eqref{eq:dws}, we find that the dilatation weights of the associated Lagrange terms are
\begin{align} \label{eq:dw135}
    \Delta \Bigl(E \, I_r^{(1,\,p)}\Bigr) = 2 \, (1 - r - p)\,, 
       \qquad%
    \Delta \Bigl(E \, I_r^{(3,\,p)}\Bigr) = \Delta \Bigl(E \, I_r^{(5,\,p)}\Bigr) = 2 \, (3 - r - p)\,.
\end{align}
Recall that $0 \leq p \leq \text{min} \{i\,, 2\}$ in $I_r^{(i,\,p)}$. In order for these terms to qualify as invariants in the NL IIB supergravity action, their associated dilatation weights have to vanish. Moreover, as we have mentioned at the end of section~\ref{sec:cqi}, components in higher-dimensional $\mathbf{S}_N^{(i)}$ vectors may be required for the full classification of all the invariants. Setting the dilatation weights in Eq.~\eqref{eq:dw135} to zero, our classification of quadratic invariants in Eq.~\eqref{eq:if2n} says that the relevant SL($2\,,\mathbb{R}$) invariant terms are (without the measure $E$)
\begin{subequations} \label{eq:invs}
\begin{align}
    I_0^{(1,\,1)} & = s^A_0 \, s^B_0 \, \eta^{}_{AB}\,, \\[4pt]
    I_1^{(3,\,2)} & = - \tfrac{1}{2} \, s^{A'AB}_1 \, s_1^{A'}{}_{\!AB} + s^{A'AB}_0 \, s^{A'}
    _2{}_{\!AB}\,,  \\[4pt]
    I_2^{(3,\,1)} & = \tfrac{1}{2} \, s^{A'B'A}_2 \, s^{A'B'}
    _2{}_{\!A} - s^{A'B'A}_1 \, s^{A'B'}
    _3{}_{\!A} + s^{A'B'A}_0 \, s^{A'B'}
    _4{}_{\!A}\,, \\[4pt]
    I_3^{(3,\,0)} & = \tfrac{1}{6} \Bigl( - s_3^{A'B'C'} s_3^{A'B'C'} + 2 s_2^{A'B'C'} s_4^{A'B'C'} - 2 s_1^{A'B'C'} s_5^{A'B'C'} + 2 s_0^{A'B'C'} s_6^{A'B'C'} \Bigr), \\[4pt]
    I_1^{(5,\,2)} & = \tfrac{1}{12} \Bigl( - s_1^{A'B'C'AB} \, s_1^{A'B'C'}{}_{\!AB} + 2 \, s_0^{A'B'C'AB} \, s_2^{A'B'C'}{}_{\!AB} \Bigr)\,, \\[4pt]
    I_2^{(5,\,1)} & = \tfrac{1}{4!} \Bigl( s_2^{A'B'C'D'A} s_2^{A'B'C'D'}{}_{\!\!A} - 2 s_1^{A'B'C'D'A} s_3^{A'B'C'D'}{}_{\!\!A} + 2 s_0^{A'B'C'D'A} s_4^{A'B'C'D'}{}_{\!\!A} \Bigr)\,.
\end{align}
\end{subequations}
Note that we have included $I_3^{(3,\,0)}$ and $I_2^{(5,\,1)}$\,, which contain $s_{5}^{(3)}$, $s_{6}^{(3)}$, $s_{3}^{(5)}$\,, and $s_4^{(5)}$ that do not appear in the polynomial realizations in Eq.~\eqref{eq:s253} and, therefore, they are not included in the NL IIB supergravity. However, according to the mapping \eqref{eq:s253}, we have
\begin{subequations}
\begin{align}
    s_0^{(3)} & = \CF^{(1)} \wedge \ell^{(2)}\,, 
        &%
    s_0^{(5)} & = \CF^{(3)} \wedge \ell^{(2)}\,, \\[4pt]
    s_1^{(3)} & = \Gamma^{(3)} = d\ell^{(2)} - \tfrac{3}{2} \, d\Phi \wedge \ell^{(2)}\,, 
        &%
    s_1^{(5)} & = \CH^{(3)} \wedge \ell^{(2)}\,,
\end{align}
\end{subequations}
and using the identities $\ell^{}_{AA'} = \ell^{}_{A'B'} = \bigl(d\ell \bigr)^{}_{A'B'C'} = 0$\,, we find 
\be \label{eq:accvanish}
    s^{A'B'C'}_{0} = s^{A'B'C'}_{1} = s^{A'B'C'D'A}_{0} = s^{A'B'C'D'A}_{1} = 0
\ee
in Eq.~\eqref{eq:s253}, which implies that the dependencies on $s_{5}^{(3)}$, $s_{6}^{(3)}$, $s_{3}^{(5)}$\,, and $s_4^{(5)}$ drop off in Eq.~\eqref{eq:invs}. Finally, plugging the mapping \eqref{eq:s253} into Eq.~\eqref{eq:invs}, we derive the list of SL($2\,,\mathbb{R}$) and dilatation invariants in Eq.~\eqref{eq:sldinvs}.

In order to identify the Chern-Simons term, we formally treat our 10D NL IIB supergravity as the boundary of an 11D theory. There exists only one zero-dilatation weight quantity $\CF^{(5)} \wedge \CF^{(3)} \wedge \CH^{(3)}$ in 11D that is invariant under SL($2\,,\mathbb{R}$) up to an exact form, such that
\be \label{eq:gf5f3h3}
    g \circ \bigl( \CF^{(5)} \wedge \CF^{(3)} \wedge \CH^{(3)} \bigr) = \CF^{(5)} \wedge \CF^{(3)} \wedge \CH^{(3)} - d \Bigl( \CF^{(5)} \wedge \CK^{(3)}_4 \wedge \ell^{(2)} \Bigr)\,.
\ee
Here,
\be
    \CK^{(3)}_4 = - \kappa \, \CH^{(3)} + \tfrac{1}{2} \, \kappa^2 \, \CF^{(3)} - \tfrac{1}{3!} \, \kappa^3 \, \Gamma^{(3)} + \tfrac{1}{4!} \, \kappa^4 \, \CF^{(1)} \wedge \ell^{(2)}
\ee
satisfies the consistency conditions \eqref{eq:consconds}. Furthermore, note that
\be \label{eq:f5f3h3}
    \CF^{(5)} \wedge \CF^{(3)} \wedge \CH^{(3)} = - d \Bigl( \CC^{(4)} \wedge \CH^{(3)} \wedge \CF^{(3)} \Bigr)
\ee
is a total derivative by itself. From Eqs.~\eqref{eq:gf5f3h3} and \eqref{eq:f5f3h3}, we find the following transformation of a ten-form quantity in the boundary 10D theory:
\be
    g \circ \Bigl( \CC^{(4)} \wedge \CH^{(3)} \wedge \CF^{(3)} \Bigr) = \CC^{(4)} \wedge \CH^{(3)} \wedge \CF^{(3)} + \CF^{(5)} \wedge \CK^{(3)}_4 \wedge \ell^{(2)} + d\chi^{(9)}\,.
\ee
Recall the transformation of $\CA^{(3)}$ in Eq.~\eqref{eq:a3trnsf}, \emph{i.e.},
\be
    g \circ \CA^{(3)} = \CA^{(3)} + \CK^{(3)}_4\,,
\ee
we find 
\be
    g \circ I^{(10)}_\text{CS} = I^{(10)}_\text{CS} + d \chi^{(9)}_{}\,,
\ee
where $I^{(10)}_{CS}$ is defined in Eq.~\eqref{eq:wzinv}. This concludes the construction for the  zero dilatation weight and preserves SL($2\,,\mathbb{R}$) up to an exact form $d\chi^{(9)}$\,.

\newpage

\bibliographystyle{JHEP}
\bibliography{nlspr}

\end{document}